\documentclass[
apj,
tighten,
twocolappendix,
twocolumn
]{aastex631}

\usepackage[T1]{fontenc}
\usepackage{CJKutf8}
\usepackage{hyperref}
\usepackage{amsmath} 
\usepackage{physics}
\usepackage{amssymb}
\usepackage{enumerate}
\usepackage{bm}
\usepackage{multirow}

\newcommand{\UCB}{\affiliation{Department of Physics, University of California, Berkeley, Berkeley, CA 94720, USA}}
\newcommand{\UNH}{\affiliation{Department of Physics \& Astronomy, University of New Hampshire, 9 Library Way, Durham NH 03824, USA}}
\newcommand{\WSU}{\affiliation{Department of Physics \& Astronomy, Washington State University, Pullman, Washington 99164, USA}}
\newcommand{\KU}{\affiliation{Department of Physics and Astronomy, KU Leuven, Celestijnenlaan 200D, B-3001 Leuven, Belgium}}

\shorttitle{
Spectral and grey neutrino transport schemes in HMNS
}
\shortauthors{Cheong et al.}
\graphicspath{{./}{figures/}}

\begin{document}

\title{
Energy-dependent and energy-integrated two-moment general-relativistic neutrino transport simulations of hypermassive neutron star 
}

\author[0000-0003-1449-3363]{Patrick Chi-Kit Cheong \begin{CJK*}{UTF8}{bkai}(張志杰)\end{CJK*}}
\email{patrick.cheong@berkeley.edu}
\altaffiliation[]{N3AS Postdoctoral fellow}
\UNH
\UCB

\author[0000-0003-4617-4738]{Francois Foucart}
\UNH

\author[0000-0002-0050-1783]{Matthew D. Duez}
\WSU

\author[0000-0002-8313-5976]{Arthur Offermans}
\KU

\author[0000-0001-8574-0523]{Nishad Muhammed}
\WSU

\author[0000-0002-3694-7138]{Pavan Chawhan}
\WSU



\begin{abstract}
We compare two-moment based \emph{energy-dependent} and 3 variants of \emph{energy-integrated} neutrino transport general-relativistic magnetohydrodynamics simulations of hypermassive neutron star.
To study the impacts due to the choice of the neutrino transport schemes, we perform simulations with the same setups and input neutrino microphysics.
We show that the main differences between energy-dependent and energy-integrated neutrino transport are found in the disk and ejecta properties, as well as in the neutrino signals. 
The properties of the disk surrounding the neutron star and the ejecta in energy-dependent transport are very different from the ones obtained using energy-integrated schemes.
Specifically, in the energy-dependent case, the disk is more neutron-rich at  early times, and becomes geometrically thicker at later times.
In addition, the ejecta is more massive, and on average more neutron-rich in the energy-dependent simulations.
Moreover, the average neutrino energies and luminosities are about 30\% higher.
Energy-dependent neutrino transport is necessary if one wants to better model the neutrino signals and matter outflows from neutron star merger remnants via numerical simulations.
\end{abstract}



\section{\label{sec:intro}Introduction}
Kilonovae, the thermal transients powered by the radioactive decay of sub-relativistic ejecta from neutron star mergers \citep{1998ApJ...507L..59L, 2010MNRAS.406.2650M, 2016AdAst2016E...8T, 2019LRR....23....1M}, are particularly interesting systems in the study of astrophysical nucleosynthesis.
Although the groundbreaking multimessenger detections of a binary neutron star merger on 17 August 2017 (e.g. \cite{2017PhRvL.119p1101A, 2017ApJ...848L..13A, 2017ApJ...848L..12A}) has confirmed our basic understanding of neutron star mergers~\citep{2017arXiv171005931M, 2018ASSL..457.....R}, details of the kilonova transients are poorly understood.
For instance, it is still unclear what is the abundance patterns of the produced elements, and thus how much neutron star mergers contribute to nucleosynthesis. 

The observables of kilonovae (i.e. their brightness, duration and colors) are sensitively related to the properties of the merger outflows (i.e. their composition, mass and velocities), which are significantly affected by neutrino transport and neutrino-matter interactions.
Neutron-rich outflows are responsible for the production of heavy elements (i.e. lanthanides and actinides) via the $r$-process, while less neutron-rich outflows produce lighter elements.
In either case, the brightness, duration and colors of the resulting light curves depends on the effective diffusion time through the ejecta, and hence on their composition, mass, and velocities \citep{2013ApJ...775...18B}.
To better connect the observational signatures of kilonovae to the matter outflows from mergers, numerical simulations of neutron star mergers with neutrino transport are necessary.

Variants of energy-integrated two-moment schemes based on the truncated moment formalism \citep{1981MNRAS.194..439T, 2011PThPh.125.1255S, 2013PhRvD..87j3004C} have been implemented in multiple neutron star merger simulation codes (e.g. \cite{2014ApJ...789L..39W, 2015PhRvD..91f4059S, 2015PhRvD..91l4021F, 2016PhRvD..94l3016F, 2018MNRAS.475.4186F, 2022MNRAS.512.1499R, 2022PhRvD.105j4028S, 2024PhRvD.109d4012S, 2024PhRvD.109d3044I, 2024MNRAS.528.5952M}) because of the good balance between accuracy and computational cost of these approximate transport schemes.
Despite its success in neutron star merger simulations, energy-integrated transport has two major limitations.
First, an ``energy closure'' has to be imposed.
Namely, one has to assume or estimate neutrino distributions as the neutrino spectrum cannot be reconstructed from the evolved moments.
This could be problematic in regions where the spectrum is non-trivial.
Second, neutrino-matter interactions that require detailed spectral information, such as neutrino-lepton inelastic scatterings and pair processes, cannot be included easily.
Either of these issues can significantly impact the evolution of neutrinos, and hence the merger dynamics and the composition of the post-merger remnant.
However, there are no study in the literature of the impact of using an energy-integrated transport algorithm in the merger context.

The main goal of this work is to better understand the importance of the evolutions of the neutrino spectrum in the evolution of post-merger hypermassive neutron stars.
To this end, we perform an energy-dependent and 3 variants of energy-integrated two-moment neutrino transport general-relativistic magnetohydrodynamics simulations of a post-merger-like hypermassive neutron star.
To focus solely on the impact of the various two-moment neutrino transport schemes on the results, we perform all simulations with identical neutrino microphysics and numerical setup, and with the same code.


The paper is organised as follows.
In section~\ref{sec:methods} we outline the methods we used in this work.
The results are presented in section~\ref{sec:results}.
This paper ends with a discussion in section~\ref{sec:discussion}.
Unless explicitly stated, we work in geometrized Heaviside-Lorentz units, for which the speed of light $c$, gravitational constant $G$, solar mass $M_{\odot}$, vacuum permittivity $\epsilon_0$ and vacuum permeability $\mu_0$ are all equal to one ($c=G=M_{\odot}=\epsilon_0 = \mu_0 = 1$).

\section{\label{sec:methods}Methods}

We use the quasi-equilibrium post-merger-like hypermassive neutron star constructed in \cite{2024arXiv240218529C} as the initial profile for our simulations.
Specifically, we pick the post-merger-like model with central energy density $\epsilon_{c}/c^2 = 1.2604\times10^{15}~\rm{g \cdot cm^{-3}}$ and angular momentum $J = 5~G M_{\odot}^2 / c$.
This equilibrium model is constructed with the ``DD2'' equation-of-state~\citep{2010NuPhA.837..210H} with a constant entropy-per-baryon $s = 1 ~ k_{\rm{B}} / \text{baryon}$ and in neutrinoless $\beta$-equilibrium.
The star is differentially rotating, following the 4-parameter rotation law of \cite{2019PhRvD.100l3019U} with angular velocity ratios $\left\{{\Omega_{\max}}/{\Omega_{\rm c}}=1.6,{\Omega_{\rm eq}}/{\Omega_{\rm c}}=1\right\}$, where $\Omega_{\max}$, $\Omega_{c}$, and $\Omega_{\rm eq}$ are the maximum, central, and equatorial angular velocities of the neutron star, respectively.
These ratios are chosen to empirically match the rotational profiles reported in binary neutron star merger simulations.
As reported in \cite{2024arXiv240218529C}, this profile is dynamically stable in conformally-flat spacetime.

Magnetic fields are added on top of the quasi-equilibrium neutron star profile.
In particular, we superimpose magnetic fields by adding the following vector potential in orthonormal form:
\begin{equation}
	\left(A^{\hat{r}}, A^{\hat{\theta}}, A^{\hat{\phi}}\right) = \frac{r_0^3 }{2\left(r^3+r_0^3\right)}\left(0, 0, B_{\rm pol} r \sin\theta \right),
\end{equation}
where the relation between orthonormal-basis components and coordinate-basis components can be found in Appendix~A of \cite{2020CQGra..37n5015C}.
Here, we set $r_0 = 10~{\rm km}$ and $B_{\rm pol} = 10^{15}~{\rm G}$.
This vector potential gives us purely poloidal magnetic fields with maximum strength $10^{15}~{\rm G}$.
At the beginning of the simulations, we impose a low-density magnetosphere with magnetic-to-gas pressure ratio $P_{\rm mag} / P_{\rm gas} = 10$ everywhere outside the star, following the methods of \cite{2015ApJ...806L..14P}.
This low-density gas is meant to mimic the gas surrounding a post-merger remnant.
Given the differential rotation of the remnant and the initial poloidal magnetic field, a disk will form around the neutron soon after the simulation starts as a result of magnetic winding.

The general-relativistic radiation magnetohydrodynamics code \texttt{Gmunu}~\citep{2020CQGra..37n5015C, 2021MNRAS.508.2279C, 2022ApJS..261...22C, 2023ApJS..267...38C, 2024ApJS..272....9N} is used for our simulations.
In particular, we dynamically evolve the general-relativistic magnetohydrodynamics equations and the Einstein field equations in the conformally flat approximation.
The divergence-free condition of the magnetic field is preserved by using staggered-meshed constrained transport \citep{1988ApJ...332..659E}.
All the simulations here are axisymmetric and performed in cylindrical coordinates $(R, z)$, where the computational domain covers $0 \leq R \leq 2000~{\rm km}$ and $0 \leq z \leq 2000~{\rm km}$, with the resolution {$n_R \times n_z = 128 \times 128$} and allowing 6 adaptive-mesh-refinement levels on top of that.
The finest grid size at the centre of the star is $\Delta R = \Delta z \approx 488 ~\rm{m}$.
The finite temperature equation-of-state ``DD2''~\citep{2010NuPhA.837..210H} is used for the evolutions.
Our simulations adopt the Harten, Lax and van Leer (HLL) approximated Riemann solver~\citep{harten1983upstream}, the 3rd-order reconstruction method PPM~\citep{1984JCoPh..54..174C} and the IMEXCB3a time integrator~\citep{2015JCoPh.286..172C}. 

Two-moment radiation transport schemes together with the maximum-entropy closure \citep{1978JQSRT..20..541M} are used for neutrino transport in this work.
\texttt{Gmunu} solves either energy-integrated or energy-dependent neutrino transport.
In this work, we implement and compare three different variants of the energy-integrated schemes with the energy-dependent scheme.
Specifically, we consider the following two-moment neutrino transport schemes:
\begin{enumerate}[(i)]
	\item \emph{$\langle \epsilon^F \rangle = \langle \epsilon \rangle$}:
		Essentially the same energy-integrated scheme as in \cite{2016PhRvD..94l3016F}, but the calculations of the neutrino number flux is simplified.
		In the neutrino number density flux (i.e. equation~\eqref{eq:fn}), the flux-weighted average energy of neutrinos $\langle \epsilon^F \rangle$ is chosen to be the same as the energy-weighted average energy $\langle \epsilon \rangle$.
		This simplification has been adopted in \cite{2022MNRAS.512.1499R, 2024PhRvD.109d4012S, 2024MNRAS.528.5952M};
	\item \emph{\cite{2016PhRvD..94l3016F}}:
		The energy-integrated neutrino transport schemes proposed by \cite{2016PhRvD..94l3016F};
	\item \emph{\cite{2024AnA...687A..55A}}:
		An improved version based on \cite{2016PhRvD..94l3016F} for core-collapse supernovae \citep{2024AnA...687A..55A}.
		Note that here we do not include modifications that are specifically calibrated to core-collapse supernovae.
	\item \emph{spectral}:
		The energy-dependent neutrino transport scheme described in \cite{2023ApJS..267...38C}.
\end{enumerate}
For the implementation details of energy-dependent radiation transport in \texttt{Gmunu}, we refer readers to \cite{2023ApJS..267...38C}.
The details of the additional implementations for energy-integrated radiation transport can be found in Appendix~\ref{sec:grey_m1}.

The neutrino microphysics considered here is similar to that of \cite{2020ApJ...902L..27F}, with interaction rates provided by \texttt{NuLib}~\citep{2015ApJS..219...24O}.
In particular, we consider three species of neutrinos: the electron type (anti)neutrinos $\nu_e$ and $\bar{\nu}_e$, and heavy-lepton (anti)neutrinos $\nu_x$, where the muon and tau neutrinos (i.e. $\nu_{\mu}, \bar{\nu}_{\mu}, \nu_{\tau}$, and $\bar{\nu}_{\tau}$) are grouped into $\nu_x$.
We further assume that neutrino-matter interactions can be described by an emissivity $\eta$, absorption opacity $\kappa_a$, and elastic scattering opacity $\kappa_s$, which are obtained from tabulated rates generated via \texttt{NuLib}.
The table includes reaction rates for the charged current reactions $p + e^- \leftrightarrow n+\nu_e$ and $n+e^+ \leftrightarrow p+\bar\nu_e$;
scattering of neutrinos on protons, neutrons, $\alpha$-particles and heavy nuclei; and $e^+e^- \leftrightarrow \nu\bar \nu$ and Bremsstrahlung for the heavy-lepton neutrinos only. 
The table is logarithmically spaced in neutrino energies $\epsilon$ (16 groups up to $528\,{\rm MeV}$), rest-mass density $\rho$ (82 points in $[10^6,3.2\times 10^{15}]\,{\rm g/cm^3}$) and fluid temperature $T$ (65 points in $[0.05,150]\,{\rm MeV}$), and linearly spaced in the electron fraction $Y_e$ (51 points in $[0.01,0.6]$). 

\section{\label{sec:results}Results}
\begin{figure}
	\centering
	\includegraphics[width=\columnwidth, angle=0]{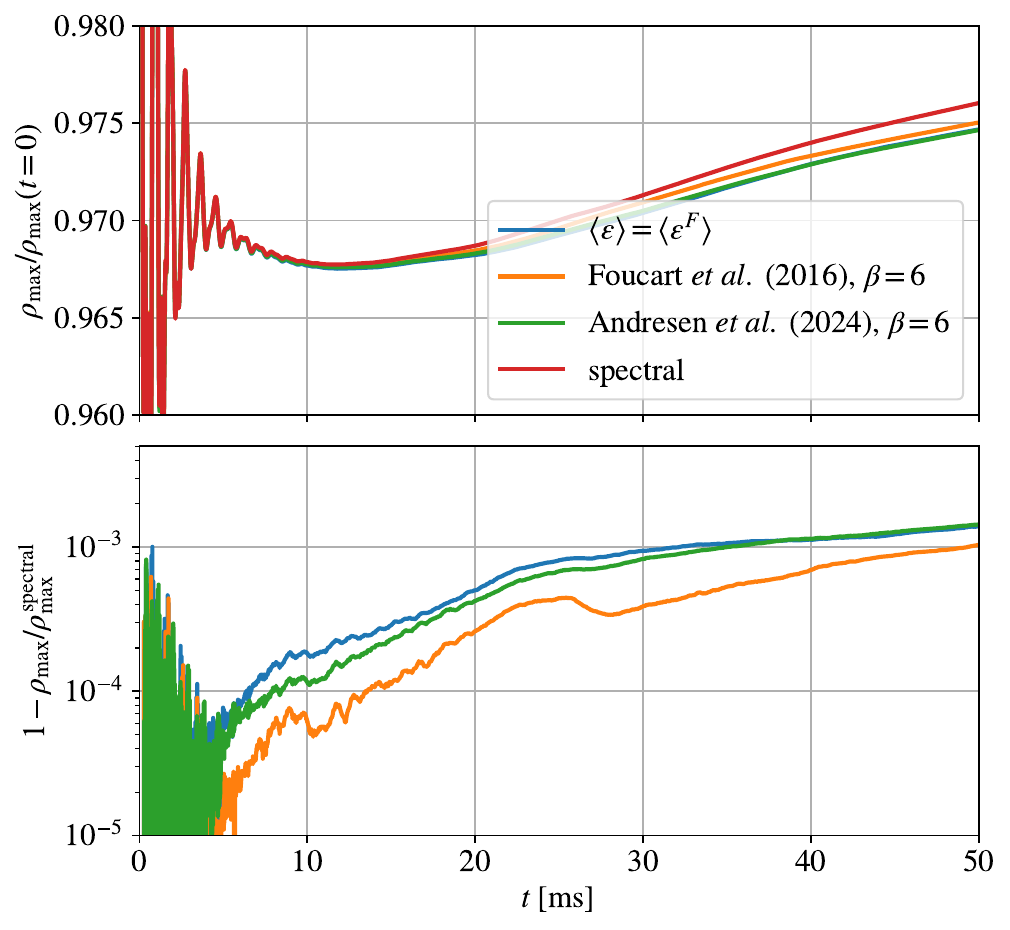}
	\caption{
		Maximum value of the rest mass density rescaled with its inital value $\rho_{\max}/\rho_{\max}\left(t=0\right)$ (\emph{top panel}) and their relative error with respect to the energy-dependent transport case (\emph{bottom panel}) as functions of time with different two-moment neutrino transport schemes of a post-merger-like hypermassive neutron star.
		In the high density regions, all the schemes work similarly.
		}
	\label{fig:ns_rho_max}	
\end{figure}

The magnetohydrodynamical evolutions of the neutron star (i.e. for $\rho \gtrsim 10^{11}~{\rm g/cm^3}$) are very similar in all cases (see, e.g. figure~\ref{fig:ns_rho_max}).
This is expected, as the neutrinos are mostly in equilibrium with the fluid in the high density and hot regions.
However, this is not the case in the matter outflows and disk (i.e. lower density regions).

\begin{figure*}
	\centering
	\includegraphics[width=\textwidth, angle=0]{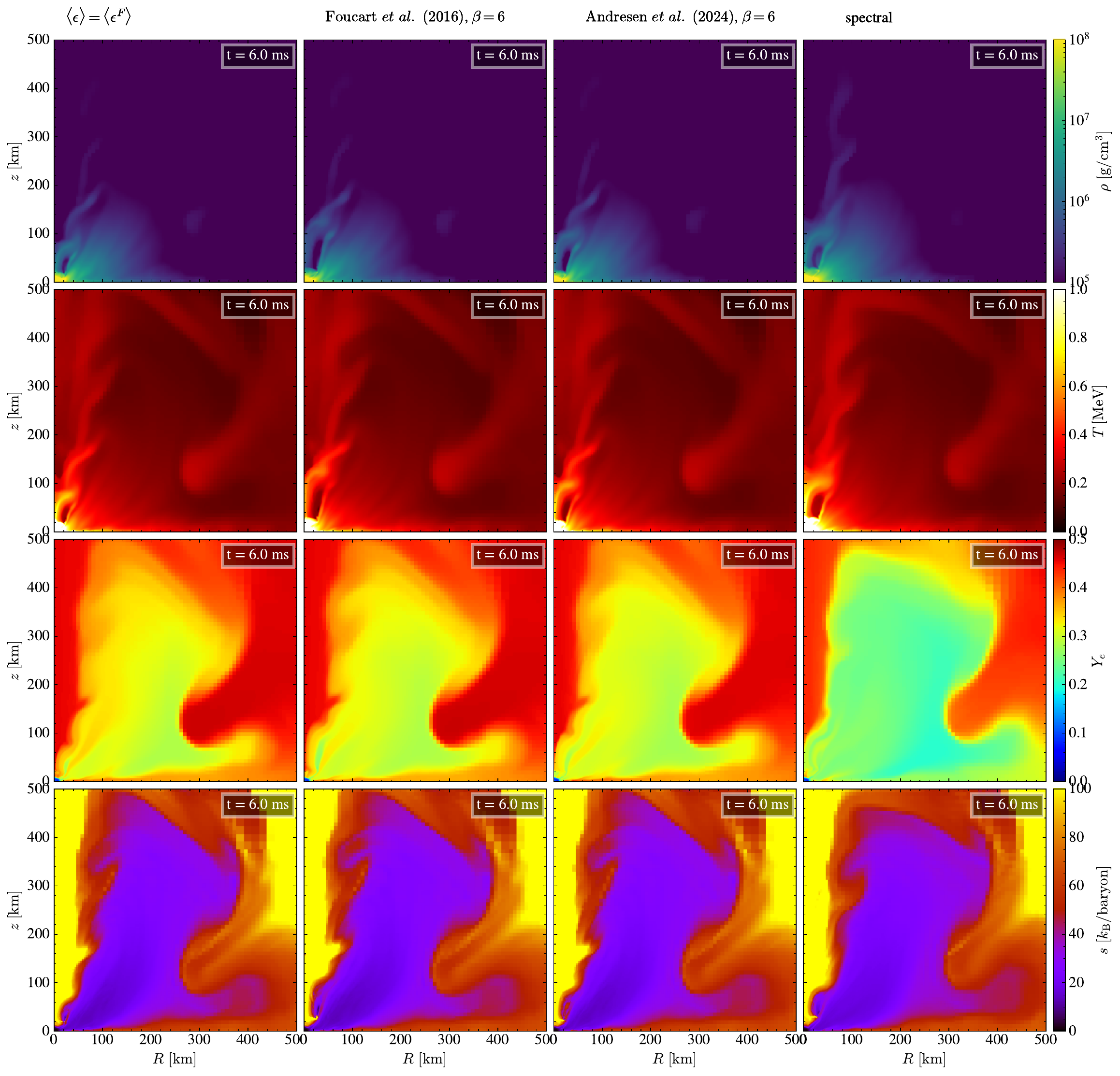}
	\caption{
		Profiles of rest-mass density $\rho$ (\emph{top row}), temperature $T$ (\emph{middle up row}), electron fraction $Y_e$ (\emph{middle down row}), entropy per baryon $s$ (\emph{bottom row}) with different two-moment neutrino transport schemes (\emph{left to right columns}) at $t=6~{\rm ms}$ of a post-merger-like hypermassive neutron star.
		Despite similar density, temperature, and entropy profiles, the low density gas is significantly more neutron-rich in the energy-dependent case.
		One of the reason for this is that the radial velocity of the low density gas flowing from the star at the beginning of the simulation, mostly due to magnetic winding, is very high. Hence this results in a large Doppler shift of the neutrino spectrum.
		This effect can only be captured when an energy-dependent scheme is used.
		Therefore, this signature appears only in the energy-dependent case.
		}
	\label{fig:ns_hydro_6ms}	
\end{figure*}

\begin{figure*}
	\centering
	\includegraphics[width=\textwidth, angle=0]{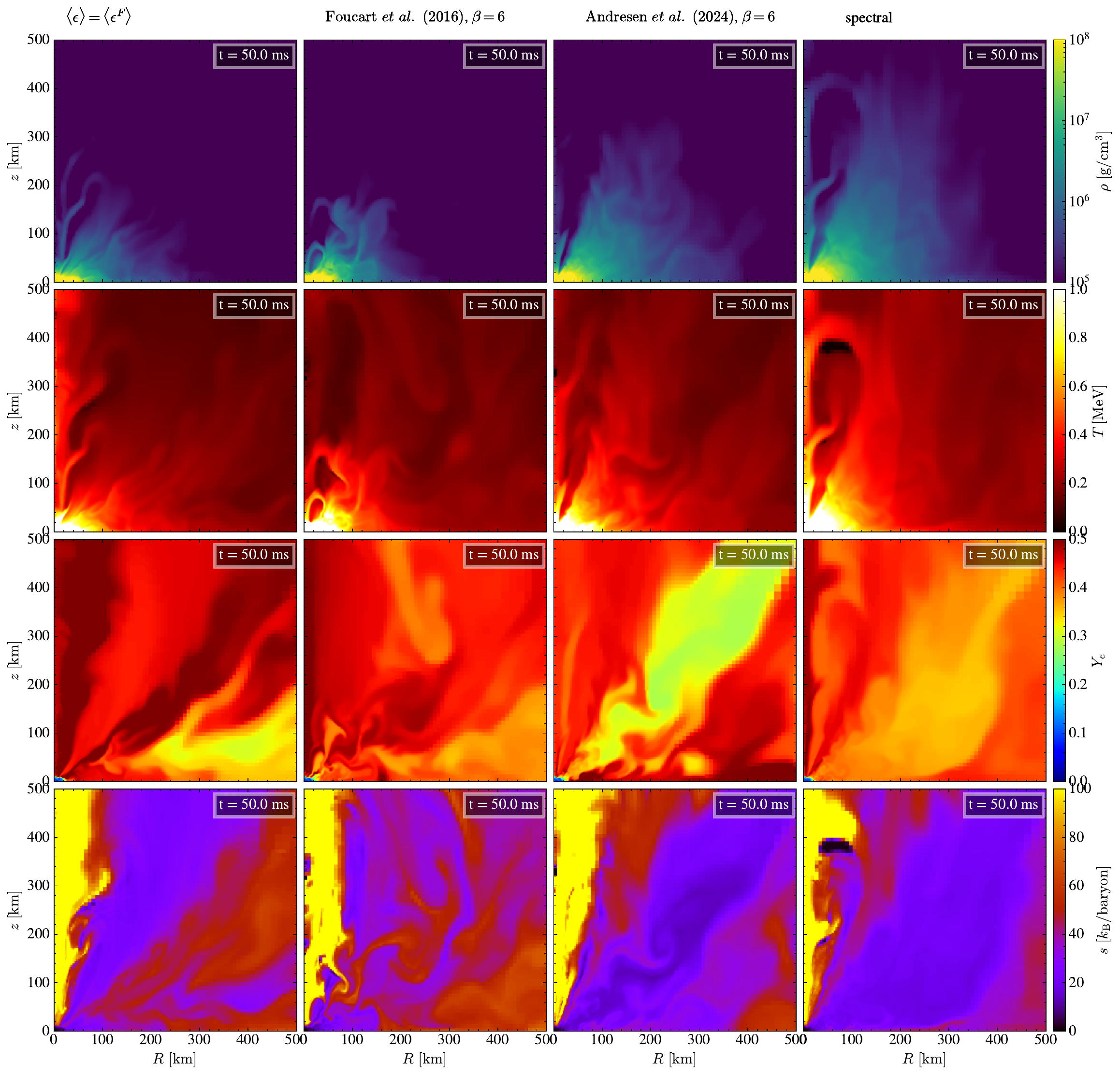}
	\caption{
		Profiles of rest-mass density $\rho$ (\emph{top row}), temperature $T$ (\emph{middle up row}), electron fraction $Y_e$ (\emph{middle down row}), entropy per baryon $s$ (\emph{bottom row}) with different two-moment neutrino transport schemes (\emph{left to right columns}) at $t=50~{\rm ms}$ of a post-merger-like hypermassive neutron star.
		At later times, the magnetic winding is mostly saturated, and a quasi-equilibrium neutron star with disk system has formed.
		The thickness and composition of the disk depends on the choice of the neutrino transport scheme.
		The disk is thicker when a more accurate moment scheme is used.
		}
	\label{fig:ns_hydro_50ms}	
\end{figure*}

To compare the properties of the disk surrounding the neutron star, we compare the rest-mass density $\rho$, temperature $T$, electron fraction $Y_e$, and entropy-per-baryon $s$ profiles at different times with different neutrino transport schemes.
At the beginning of the simulations, low density gas flow out with non-negligible radial velocity from the surface of the star due to the magnetic winding effect.
The Doppler shift effect significantly alters the distribution of the neutrinos in those regions, which can only be captured when an energy-dependent transport scheme is used.
As a result, at early times, although the density, temperature, and entropy are qualitatively the same, the electron fraction is distinctively lower in the energy-dependent case, as shown in figure~\ref{fig:ns_hydro_6ms}.
At later times, the magnetic winding effect mostly saturates and the system reaches a quasi-equilibrium state, resulting in a neutron star with disk system.
As shown in the hydrodynamical profiles at $t=50~{\rm ms}$ in figure~\ref{fig:ns_hydro_50ms}, the more accurate the neutrino transport algorithm is, the thicker the disk becomes.

\begin{figure}
	\centering
	\includegraphics[width=\columnwidth, angle=0]{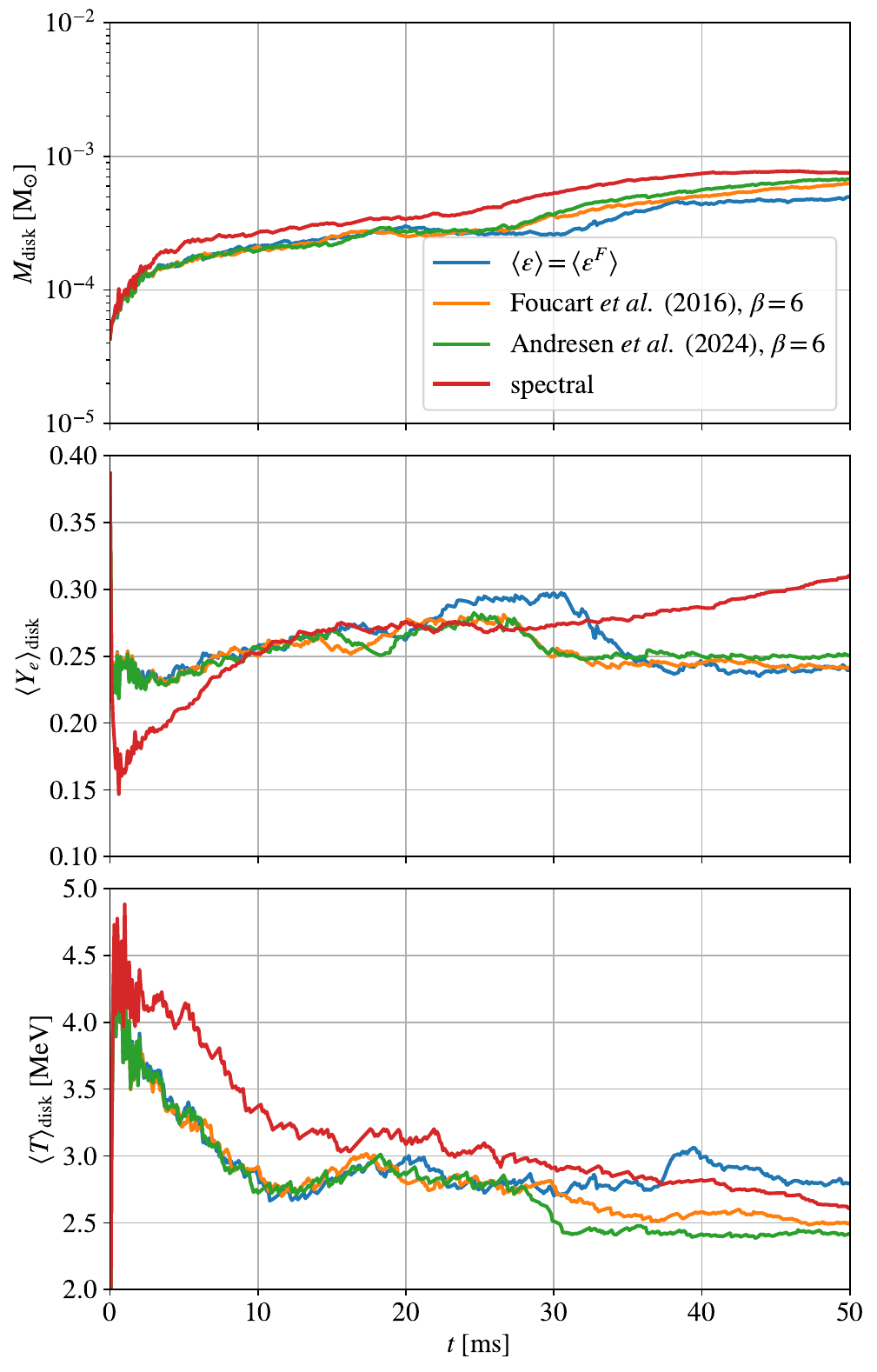}
	\caption{
		The total rest-mass, mass-averaged electron fraction, temperature of the disk as functions of time with different two-moment neutrino transport schemes of a post-merger-like hypermassive neutron star.
		Here, we define the disk as the regions where $\rho \leq 10^{11}~{\rm g \cdot cm^{-3}}$.
		In the energy-dependent case, disk mass is slightly higher.
		Initially the disk is more neutron-rich, but become more neutron-poor at later time.
		}
	\label{fig:ns_disk_avg}
\end{figure}

Figure~\ref{fig:ns_disk_avg} shows the disk mass and the density-weighted averaged electron fraction and temperature.
Those quantities are defined as
\begin{align}
	{M}_{\rm disk} = & \int_{\rho \leq 10^{11}~{\rm g \cdot cm^{-3}}} D \dd{V} , \\
	\langle {X} \rangle_{\rm{disk}} = & \frac{1}{{M}_{\rm disk}}\left[ \int_{\rho \leq 10^{11}~{\rm g \cdot cm^{-3}}} D X \dd{V} \right], 
\end{align}
where the variables $X$ here can be either temperature $T$ or electron fraction $Y_e$.
In the energy-dependent case, the disk mass is slightly higher.
Initially the disk is more neutron-rich, but become more neutron-poor at later time.

Detailed properties of the ejecta, such as its mass, composition, geometry, and entropy, are very important to observables of kilonovae.
Here, the matter is identified as unbound when it fullfills the Bernoulli criteria and is outgoing. 
In particular, we locate the unbound matter everywhere in the computational domain by checking
\begin{align}\label{eq:unbound}
	\begin{cases}
		& h u_t \leq -h_{\min} \\
		& v_r > 0
	\end{cases},
\end{align}
where $h$ is specific enthalpy while $h_{\min}$ is its minimum allowed values for a given equation-of-state, $u_t = W(- \alpha + \beta_i v^i )$, and $v_r$ is the radial velocity.

\begin{figure*}
	\centering
	\includegraphics[width=\textwidth, angle=0]{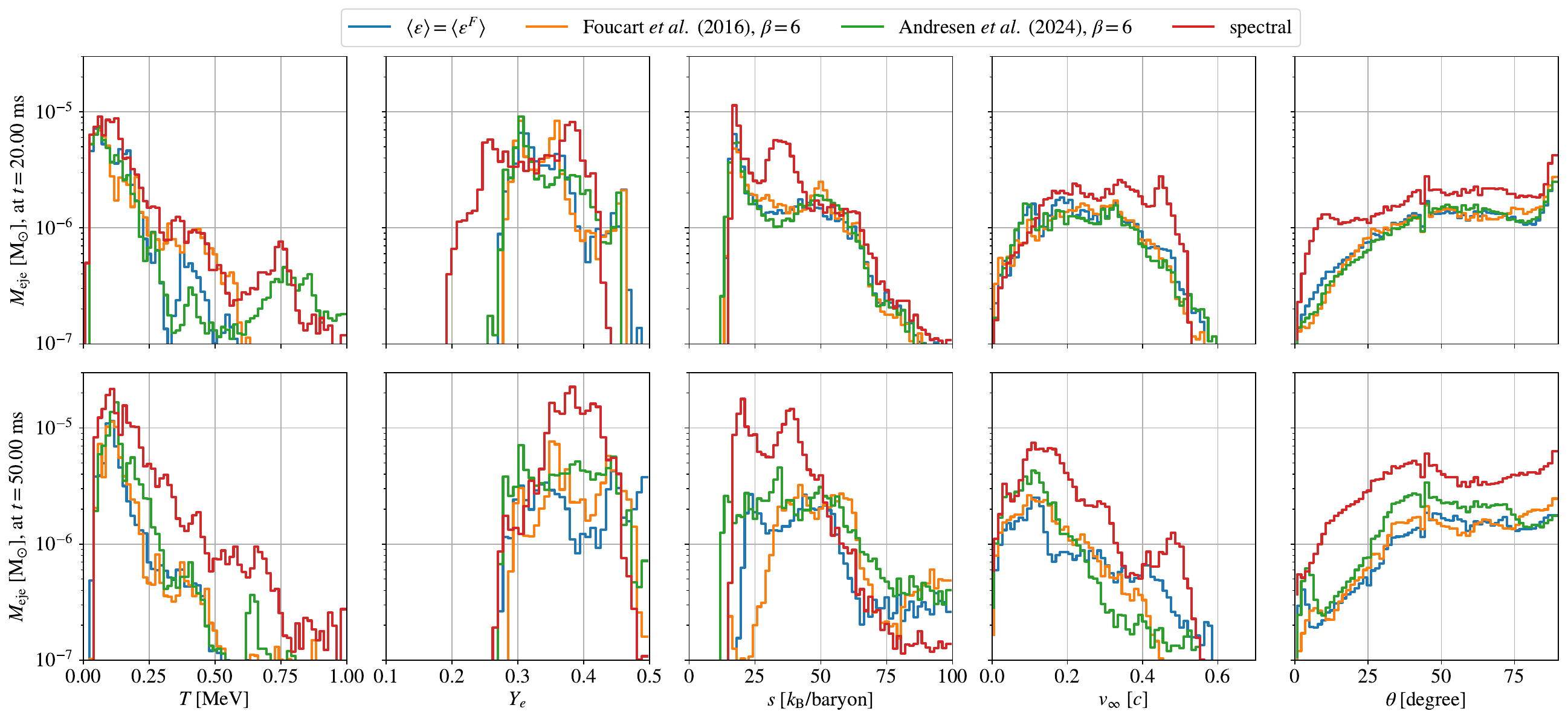}
	\caption{
		1-D histograms of the unbound outgoing matter in the computational domain of a post-merger-like hypermassive neutron star at $t=20~{\rm ms}$ (\emph{upper panel}) and $t=50~{\rm ms}$ (\emph{lower panel}).
		These plots compare the distribution of ejected mass as functions of temperature $T$, electron fraction $Y_e$, entropy per baryon $s$, asymptotic velocity $v_{\infty}$ and the angle ($\theta \in \left[ 0^{\circ} , 90^{\circ} \right]$ from pole to equatorial plane) with different two-moment neutrino transport schemes.
		The ejecta behaves similarly in all the energy-integrated cases (blue, orange, and green lines), but not in the energy-dependent case (red lines).
		At early times, simulations with energy-dependent neutrino transport predict more neutron-rich and faster matter outflows than the energy-integrated schemes do.
		Those neutron-rich ejecta leaves the computational domain gradually and no longer be captured in the histograms at later times.
		At later times, the composition of the ejecta in the computational domain become less neutron-rich in the energy-dependent scheme.
		These results suggest that the ejecta properties sensitively depend on the accuracy of the spectral information of neutrinos.
		}
	\label{fig:ns_eje_histogram}
\end{figure*}

Figure~\ref{fig:ns_eje_histogram} compares the ejecta properties for the different two-moment neutrino transport schemes at $t=20~{\rm ms}$ and $t=50~{\rm ms}$.
All the energy-integrated neutrino transport schemes predict very similar ejecta properties.
However, the ejecta behaves differently when an energy-dependent neutrino transport is used.
The total amount of ejecta is larger in the energy-dependent case, in all direction.
Moreover, the distributions of the electron fraction $Y_e$, the entropy per baryon $s$ and the asymptotic velocity $v_{\infty}$ are noticeably different from the energy-integrated cases.
The energy-dependent scheme predicts more neutron-rich matter in the equatorial plane, and faster matter outflows in the polar regions.
This implies that energy-integrated schemes could potentially underestimate the total amount of neutron-rich and/or fast ejecta in neutron star mergers.

We further compare the total mass and the density-weighted average properties of the ejecta.
The total mass ${M}_{\rm eje}$ and the density-weighted averaged variable $X$ of the ejecta can be estimated by \citep{2023CQGra..40h5008H}
\begin{align}
	{M}_{\rm eje} = &\int_0^t dt' \oint_{S_{\rm ext}} \hat{v}^i D \dd{A_i} + \int_{V_{\rm ext}} D f_{\rm ub} \dd{V} , \label{eq:matter_flux} \\
	\langle {X} \rangle_{\rm{eje}} = & \frac{1}{{M}_{\rm eje}}\left[ \int_0^t dt' \oint_{S_{\rm ext}} \hat{v}^i D X \dd{A_i} + \int_{V_{\rm ext}} D X f_{\rm ub} \dd{V} \right], \label{eq:avg_flux}
\end{align}
where $\hat{v}^i = \alpha v^i - \beta^i$, $D=W\rho$ is the conserved rest-mass density, $f_{\rm ub}=1$ when the fluid at a point is unbound (i.e. fullfiling equation~\eqref{eq:unbound}) while equals to zero elsewhere, and the variables $X$ can be temperature $T$, entropy $s$, electron fraction $Y_e$, etc.
$S_{\rm ext}$ is the extraction surface, which is chosen to be a cylinder with radius $R=1800~{\rm km}$ and $\lvert z \rvert =1800~{\rm km}$, while $V_{\rm ext}$ is the corresponding enclosed region.

\begin{figure}
	\centering
	\includegraphics[width=\columnwidth, angle=0]{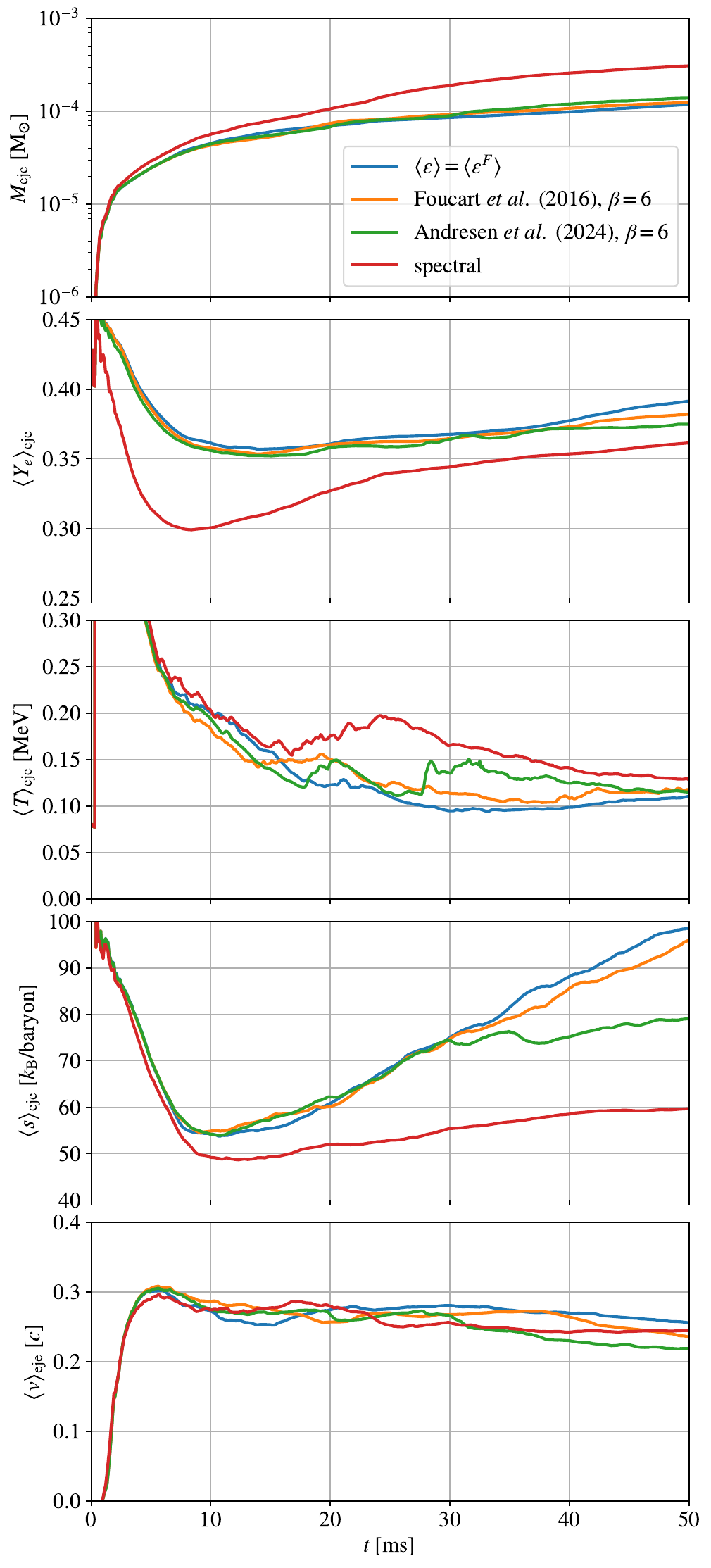}
	\caption{
		The total rest-mass, mass-averaged electron fraction, temperature, entropy, and asymptotic velocity of the ejecta as functions of time with different two-moment neutrino transport schemes of a post-merger-like hypermassive neutron star.
		In the energy-dependent case, the ejecta properties differ noticeably compared to all the energy-integrated cases.
		At $t \approx 50~{\rm ms}$, the ejected mass is about 2 times higher.
		In addition, the ejecta is overall more neutron-rich, and has lower entropy.
		This implies that energy-integrated schemes could underestimate neutron-rich mass ejection in neutron star mergers.
		}
	\label{fig:ns_eje_avg}
\end{figure}

Figure~\ref{fig:ns_eje_avg} compares the total rest-mass, mass-averaged electron fraction, temperature, entropy, and asymptotic velocity of the ejecta as functions of time with different two-moment neutrino transport schemes of a post-merger-like hypermassive neutron star.
All the energy-integrated schemes result in minor differences in mass ejection, but are very different from the energy-dependent case.
In the energy-dependent case, the ejected mass is about 2 times higher by $t \approx 50~{\rm ms}$.
In addition, the ejecta is overall more neutron-rich, and has lower entropy.
This implies that energy-integrated schemes could underestimate neutron-rich mass ejection in neutron star mergers.

\begin{figure*}
	\centering
	\includegraphics[width=\textwidth, angle=0]{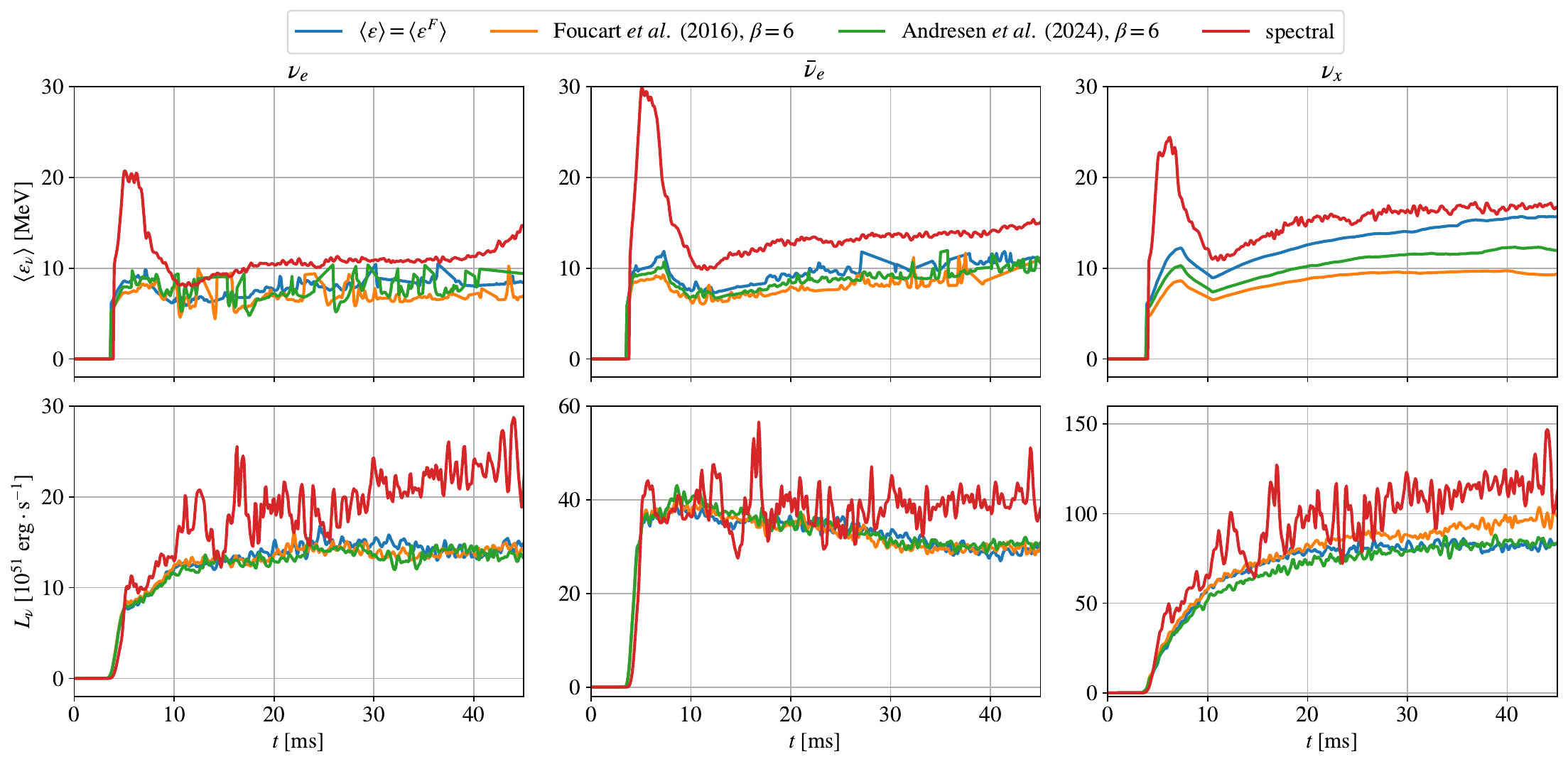}
	\caption{
	Time evolution of far-field averaged neutrino energies ($\langle {\epsilon_\nu} \rangle$, \emph{upper panels}) and luminosities ($L_\nu$, \emph{lower panels}) measured by an observer comoving with fluid at 1000 km of a post-merger-like hypermassive neutron star.
	The averaged neutrino energies are qualitatively the same among all the cases.
	The energy hierarchy $\langle \epsilon_{\nu_x} \rangle > \langle \epsilon_{\bar{\nu}_e} \rangle > \langle \epsilon_{\nu_e} \rangle$ is observed among all neutrino transport schemes we adopted here, which agrees with the literature.
	Similarly, the luminosities in all cases agree qualitatively.
		}
	\label{fig:ns_nu}	
\end{figure*}

Finally, we present the neutrino signals from the system. 
Figure~\ref{fig:ns_nu} compares the time evolution of the average energies of neutrinos $\langle {\epsilon_\nu} \rangle$ and luminosities $L_\nu$ at 1000 km of a post-merger-like hypermassive neutron star.
Here, the neutrino luminosities $L_\nu$ observed by an observer are obtained using
\begin{equation}
	L_{\nu} = \oint \alpha F^i_\nu - \beta^i E_\nu \dd{A_i}, \label{eq:nu_flux}
\end{equation}
where $\dd{A_1} = \sqrt{{\gamma}} dx^{2} dx^{3}$ (and $\dd A_{2,3}$ are obtained by permutation of the indices).
The surface of this integral is chosen to be a cylinder with radius $R=1000~{\rm km}$ and $\lvert z \rvert =1000~{\rm km}$.
The average energies and luminosities of the energy-dependent scheme are about 30\% higher than in the energy-integrated schemes.
For the energy-integrated cases, the average neutrino energy of electron type neutrinos $\langle \epsilon_{\nu_{e}}\rangle$ and $\langle \epsilon_{\bar{\nu}_{e}}\rangle$ are very close for all cases.
However, the average energy of the heavy-lepton neutrinos have visible differences between energy-integrated schemes.
This is because the absorption neutrinosphere is deeper into the star than the scattering neutrinosphere for heavy-lepton neutrinos, and the treatment of the energy spectrum in the regions between those two neutrinospheres is the major difference between all the schemes we include here.
Note that, although our initial condition is constructed to rotate similarly to those reported from direct neutron star merger simulations, the thermodynamics properties are very different at the stellar centre.
Typically, the central temperature of a post-merger hypermassive neutron star is cooler than its surface, while the opposite is true in our constant-entropy initial data.
Therefore, the neutrino signals reported here, especially for heavy-lepton neutrinos which are emitted mostly from the inner part of the star, may not agree well with the results of full neutron star merger simulations.
The energy hierarchy $\langle \epsilon_{\nu_x} \rangle > \langle \epsilon_{\bar{\nu}_e} \rangle > \langle \epsilon_{\nu_e} \rangle$ (e.g. \cite{1998A&A...338..535R, 2020ApJ...902L..27F}) is observed marginally in all cases except the case using the methods of \cite{2016PhRvD..94l3016F}, where $\langle \epsilon_{\nu_x}\rangle \lesssim \langle \epsilon_{\bar{\nu}_e} \rangle$.

\section{\label{sec:discussion}Discussion}
In this work, we compare two-moment energy-dependent and energy-integrated neutrino transport simulations of hypermassive neutron stars.
Specifically, we perform general-relativistic radiation magnetohydrodynamics simulations of high angular momentum hypermassive neutron star with an energy-dependent and 3 variants of energy-integrated two-moment schemes, with the same input neutrino microphysics.
We study the impact of this choice of the neutrino transport schemes on our simulations, focusing particularly on the neutrino signatures as well as the disk and ejecta properties.

As expected, we find that all the schemes work similarly in the high density hot regions, as the neutrinos are mostly in equilibrium with the fluid.
We do not find any significant difference in the evolution of the neutron star among all 4 different neutrino transport schemes.
However, this is not the case for the semi-transparent region where the spectral information of the neutrinos is essential yet non-trivial.
Not only are the neutrinos not fully in equilibrium with the fluid, but also their spectrum can be significantly altered by the gravitational redshift and the Doppler shift effects.

The properties of the disk surrounding the star are significantly different when an energy-dependent scheme is used. 
Dependents on the characteristics of the magnetic winding effects, either the gravitational redshift or the Doppler shift effects can impact the neutrino spectrum at different level.
These effects can only be captured with energy-dependent transports, and hence lead to very different disk formation process compared to energy-integrated simulations.
Compared to the energy-integrated cases, we find that the disk in the energy-dependent case is more neutron-rich at early times ($t \lesssim 10~{\rm ms}$), and becomes geometrically thicker at later times ($t \gtrsim 50~{\rm ms}$).

The evolution of the matter outflows (i.e. their thermodynamical properties and composition), which are sensitively dependents on neutrino-matter interactions, are also heavily dependent on the accuracy of the neutrino energy distributions.
In our simulations, we find that the properties of the ejecta are noticeably different between energy-dependent and energy-integrated schemes. 
In the energy-dependent case, at early times, it gives more neutron-rich and faster matter outflows than the energy-integrated schemes, while the composition of the outflows are more similar at later times.
The total ejecta is overall more massive, more neutron-rich, and has lower entropy in the energy-dependent simulation.

The energy-dependent scheme predicts average energies and luminosities $\sim 30\%$ higher  than the energy-integrated cases.
The difference in average neutrino energies between the different energy-integrated schemes are on the other hand very small, except for heavy-lepton neutrinos. 
This is because different schemes have different treatment of the regions between the absorption and scattering neutrinosphere, and these regions are larger for heavy-lepton neutrinos than for electron types neutrinos.

Our findings agree with the comparison of (energy-dependent) Monte-Carlo and energy-integrated two-moment schemes of direct neutron star merger simulations presented in \cite{2020ApJ...902L..27F, 2024arXiv240715989F}.
For instance, \cite{2020ApJ...902L..27F, 2024arXiv240715989F} have also reported higher neutrino average energies and luminosities in their Monte-Carlo simulations than in their energy-integrated two-moment simulation, as well as higher ejected masses (although the latter is likely not resolved at the current accuracy of the Monte-Carlo simulations).
Both their and our results suggest that estimating the neutrino distributions based on \cite{2016PhRvD..94l3016F, 2024arXiv240715989F} in energy-integrated transports could underestimate the neutrino average energies and luminosities, as well as the ejected mass.

The uncertainties due to neutrino microphysics could be even larger than the discrepancy between energy-dependent and energy-integrated transports.
Most of the neutron star merger simulations consider only the iso-energetic neutrino-matter interactions. More complicated neutrino-matter interactions that involve energy or species coupling such as neutrino-lepton inelastic scattering and pair processes are often ignored, or treated very approximately.
These neutrino-matter interactions could be important in neutron star mergers \citep{2024ApJS..272....9N}, altering the evolution of the entire system.
Spectral information about neutrinos is needed in order to include these interactions in simulations properly.
The inclusion of these interactions in hypermassive neutron star simulations is left as future work.

Energy-dependent neutrino transport is essential for accurate kilonova modelling.
The observables of kilonovae are sensitively related to the properties of the merger outflows. 
Here, we show that the ejecta properties are sensitive to the neutrino spectral information.
The timescales of these effects and differences lie into the typical lifetime of a post-merger hypermassive neutron star ($\mathcal{O}\left(10~{\rm ms}\right)$).
Depending on when the hypermassive neutron star collapses to a black hole, the resulting disk and ejecta could be very inaccurate if energy-integrated schemes are used.
Moreover, the ability to resolve the neutron spectral information is essential to include more neutrino microphysics.
To accurately model matter outflow from neutron star mergers, energy-dependent neutrino transport is necessary.

\begin{acknowledgments}
P.C.K.C. thanks Haakon Andresen and Evan O’Connor for providing the \texttt{NuLib} table and the proto-neutron star profile for the test presented in Appendix~\ref{sec:pns_test}, and the useful discussions about this project. 
P.C.K.C. also thanks David Radice, Carlos Palenzuela and Stephan Rosswog for useful discussions over the duration of this project. 
P.C.K.C. gratefully acknowledges support from NSF Grant PHY-2020275 (Network for Neutrinos, Nuclear Astrophysics, and Symmetries (N3AS)).
F.F. gratefully acknowledge support from the Department of Energy, Office of Science, Office of Nuclear Physics, under contract number DE-AC02-05CH11231 and from the NSF through grant AST-2107932. 
M.D. gratefully acknowledges support from the NSF through grant PHY-2110287.  
M.D. and F.F. gratefully acknowledge support from NASA through grant 80NSSC22K0719. 

The simulations in this work have been performed on the UNH supercomputer Marvin, also known as Plasma, which is supported by NSF/MRI program under grant number AGS-1919310. 
This work also used Expanse cluster at San Diego Supercomputer Centre through allocation PHY230104 and PHY230129 from the Advanced Cyberinfrastructure Coordination Ecosystem: Services \& Support (ACCESS) program~\cite{10.1145/3569951.3597559}, which is supported by National Science Foundation grants \#2138259, \#2138286, \#2138307, \#2137603, and \#2138296. 
\end{acknowledgments}

%



\software{
{
The results of this work were produced by utilising \texttt{Gmunu}~\citep{2020CQGra..37n5015C, 2021MNRAS.508.2279C, 2022ApJS..261...22C, 2023ApJS..267...38C, 2024ApJS..272....9N}, where the tabulated neutrino interaction were provided with \texttt{NuLib}~\citep{2015ApJS..219...24O}.
Conformally flat, axisymmetric, differentially rotating hot neutron stars in quasi-equilibrium are constructed using the \texttt{RotNS} code~\citep{1994ApJ...422..227C, 2024arXiv240218529C, 2024arXiv240305642M}.
The data of the simulations were post-processed and visualised with 
\texttt{yt}~\citep{2011ApJS..192....9T},
\texttt{NumPy}~\citep{harris2020array}, 
\texttt{pandas}~\citep{reback2020pandas, mckinney-proc-scipy-2010},
\texttt{SciPy}~\citep{2020SciPy-NMeth} and
\texttt{Matplotlib}~\citep{2007CSE.....9...90H, thomas_a_caswell_2023_7697899}.
}
}


\appendix

\section{\label{sec:grey_m1}Implementation of energy-integrated neutrino transport schemes}
\subsection{Moments evolutions}
The evolution equations of the moments of neutrinos can be obtained by \citep{2011PThPh.125.1255S, 2013PhRvD..87j3004C, 2020LRCA....6....4M}
\begin{equation}\label{eq:evolution}
	\nabla_\nu \mathcal{T}^{\mu\nu} - \frac{1}{\epsilon^2}\frac{\partial}{\partial \epsilon} \left( \epsilon^2 \mathcal{U}^{\mu\nu\rho} \nabla_\rho u_\nu \right) = \mathcal{S}^\mu_{\rm{rad}},
\end{equation}
where $\mathcal{T}^{\mu\nu}$ and $\mathcal{U}^{\mu\nu\rho}$ are the monochromatic second-rank and third-rank energy-momentum tensors, and $\mathcal{S}^\mu_{\rm{rad}}$ is the radiation four-force, which describes the interaction between the radiation and the fluid.
Energy-integration of equation~\eqref{eq:evolution} gives
\begin{equation}\label{eq:grey_evolution}
	\nabla_\nu {T}^{\mu\nu}_{\rm rad} = {S}^\mu_{\rm{rad}},
\end{equation}
where
\begin{align}
	T^{\mu\nu}_{\text{rad}} = & \int_0^\infty 4 \pi \epsilon^2 \mathcal{T}^{\mu\nu} \dd{\epsilon}, \\
	S^{\mu}_{\text{rad}} = & \int_0^\infty 4 \pi \epsilon^2 \mathcal{S}^{\mu}_{\rm rad} \dd{\epsilon}
\end{align}
are the energy-integrated energy-momentum tensor of the radiation and the radiation four-force.
The major limitation of energy-integrated transport is two-fold:
(i) the evolution of the neutrino spectrum cannot be captured; and
(ii) neutrino-matter interactions that require spectral information of neutrinos cannot be easily included.
Either of them can significantly impact the evolution of neutrinos, their coupling to the fluid, and hence the merger dynamics and the evolution of the fluid composition.

The advection in energy space (i.e. the second term of the left-hand-side of equation~\eqref{eq:evolution}), which captures the gravitational and Doppler red/blue-shifts of neutrino spectrum, vanishes after energy integration because of energy conservation.
In neutron star mergers, this term is not negligible as the gravitational effect is strong and the fluid velocity is high.
Properly capture the energy advection is essential to neutrino evolutions (e.g. see the energy advection tests in \cite{2015ApJS..219...24O, 2016ApJS..222...20K, 2020ApJ...900...71A, 2020MNRAS.496.2000C, 2023ApJS..267...38C}).
Without taking care of the energy advection could significantly affect neutrino evolutions, and lead to unaccurate results.

Here, we adopt the $3+1$ reference-metric formalism \citep{2014PhRvD..89h4043M, 2020PhRvD.101j4007M, 2020PhRvD.102j4001B}.
In this formalism, the metric can be written as
\begin{equation}
	ds^2 = -\alpha^2  dt^2 + \gamma_{ij} \left( dx^i + \beta^i dt \right) \left( dx^j + \beta^j dt \right),
\end{equation}
where $\alpha$ is the lapse function, $\beta^i$ is the spacelike shift vector and $\gamma_{ij}$ is the spatial metric.
We adopt a conformal decomposition of the spatial metric $\gamma_{ij}$ with the conformal factor $\psi$: $ \gamma_{ij} = \psi^4 \bar{\gamma}_{ij}$, where $\bar{\gamma}_{ij}$ is the conformally related metric.
In the conformally flat approximation, the reference metric is the conformally related metric (i.e. $\bar{\gamma}_{ij}=\hat{\gamma}_{ij}$).

The evolution equations of the first two moments of radiations $E$ and $F_i$ for each species can be written as
\begin{align}
	&\begin{aligned}\label{eq:evolution_e}
		\frac{\partial}{\partial t} \left[ \sqrt{{\gamma}/\hat{\gamma}} {E} \right] 
		& + \hat{\nabla}_i \left[\sqrt{{\gamma}/\hat{\gamma}} \left(\alpha {{F}}^i - {E} \beta^i \right)\right] \\
	 	= &\sqrt{{\gamma}/\hat{\gamma}} \left[ - {{F}}^j \partial_j \alpha  + {{P}}^{ij}K_{ij} \right] \\
		 &- \alpha \sqrt{{\gamma}/\hat{\gamma}} {S}_{\rm{rad}}^\mu n_\mu, 
	\end{aligned} \\
	&\begin{aligned}\label{eq:evolution_f}
		\frac{\partial}{\partial t} \left[ \sqrt{{\gamma}/\hat{\gamma}} {F}_i \right] 
		& + \hat{\nabla}_i \left[\sqrt{{\gamma}/\hat{\gamma}} \left( \alpha {{P}}^i_{\; j} - {{F}}_j \beta^i \right)\right] \\
	 	= &\sqrt{{\gamma}/\hat{\gamma}} \left[ - {E} \partial_i \alpha + {{F}}_k \hat{\nabla}_i \beta^k + \frac{1}{2} \alpha {{P}}^{jk} \hat{\nabla}_i \gamma_{jk}\right] \\
		& + \alpha \sqrt{{\gamma}/\hat{\gamma}} {S}_{\rm{rad}}^\mu \gamma_{i\mu}, 
	\end{aligned}
\end{align}
where $n^\mu$ is the four-velocity of an Eulerian observer, $\hat{\nabla}_i$ is the covariant derivatives associated with the reference metric $\hat{\gamma}_{ij}$, $P_{ij}$ is the radiation stress tensor, and $K_{ij}$ is the extrinsic curvature.

\subsection{Neutrino number density evolutions}
In addition to neutrino energy and flux density, we also evolve neutrino number density $N$
\begin{equation}\label{eq:evolution_n}
	\frac{\partial}{\partial t} \left[ \sqrt{{\gamma}/\hat{\gamma}} N \right] 
	+ \hat{\nabla}_i \left[\sqrt{{\gamma}/\hat{\gamma}} \left(\alpha {{F}_N}^i - {N} \beta^i \right)\right] 
	= \alpha \sqrt{{\gamma}/\hat{\gamma}} C_{0}, 
\end{equation}
where the $\hat{\nabla}_i$ here is the covariant derivatives associated with the reference metric $\hat{\gamma}_{ij}$.
As in \cite{2016PhRvD..94l3016F}, we reconstruct $N/E$ for the flux calculation in equation~\eqref{eq:evolution_n}, and apply HLL Riemann solver.

The ${F_N}^i$ in the flux terms in equation~\eqref{eq:evolution_n} is
\begin{equation}\label{eq:fn}
	{F_N}^i = \frac{JWv^i}{\langle \epsilon \rangle} + \frac{\gamma^i_\mu H^\mu}{\langle \epsilon^F \rangle},
\end{equation}
where $\langle \epsilon \rangle$ is an energy-weighted average neutrino energy, which can reasonably be estimated by
\begin{equation}\label{eq:avg_eps}
	\langle \epsilon \rangle = W \frac{E - F_i v^i}{N}.
\end{equation}
$\langle \epsilon^F \rangle$ is a flux-weighted average energy of the neutrino, the choice of which is a major limitation of this energy-integrated scheme.

One of the simplest choice is by setting $\langle \epsilon^F \rangle = \langle \epsilon \rangle$. 
This choice is very convenient from the implementation point of view, and has been adopted in \cite{2022MNRAS.512.1499R, 2024PhRvD.109d4012S, 2024MNRAS.528.5952M}.
Although $\langle \epsilon^F \rangle \approx \langle \epsilon \rangle$ in the optically thin regions, the choice of $\langle \epsilon^F \rangle$ is non-trivial elsewhere.
In low absorption optical depth but high scattering optical depth regions, the spectrum of the neutrino flux can be significantly biased towards lower neutrino energies (i.e. $\langle \epsilon^F\rangle < \langle \epsilon\rangle$).

Here, we consider the scheme proposed by \cite{2016PhRvD..94l3016F}.
Specifically, in this scheme, $\langle \epsilon^F \rangle$ is chosen to be
\begin{equation}\label{eq:eF}
	\frac{\langle \epsilon^F \rangle }{\langle \epsilon \rangle} = 
	\frac{F_{3}F_{0} - s^F\left( F_{3}F_{0}-F_{2}F_{1} \right)}{F_{3}F_{0} - s^C\left( F_{3}F_{0}-F_{2}F_{1} \right)},
\end{equation}
where $F_a=F_a\left(\frac{\mu_\nu}{k_{\rm B}T}\right)$ is the $a$-th order complete Fermi-Dirac integral, and $\mu_\nu$ is the chemical potential for neutrinos.
For a more compact expression, we omit the dependence in $\frac{\mu_\nu}{k_{\rm B}T}$ in equation~\eqref{eq:eF}.
Two auxiliary scalars $s^C$ and $s^F$, where $0<s^C<s^F<1$, are introduced to control $\langle \epsilon^F \rangle/\langle \epsilon \rangle$ as shown in equation~\eqref{eq:eF}.
$s^C$ represents the fraction of neutrino which have gone through a significant optical depth since emission.
$s^F$ is chosen to be a function of $s^C$ and of the optical depth $\tau$
\begin{equation}
	{s^F} = \frac{s^C + \tau}{1+\tau}.
\end{equation}
By following \cite{2016PhRvD..94l3016F}, the optical depth $\tau$ is estimated as 
\begin{equation}\label{eq:zeta}
	{\zeta} = 1/\left(1+\beta\tau\right),
\end{equation}
where $\zeta \equiv \sqrt{{{H}^\mu {H}_\mu}/{{J}^2}}$ is the flux factor, while $\beta$ is a parameter for this scheme (not to be confused with the shift vector).
$\beta$ is calibrated to be 4 to 8 for core-collapse supernovae system, such that the optical depth $\tau$ is roughly $2/3$ on the neutrinosphere.

Here, we dynamically evolve $s^C$ in each implicit step by
\begin{equation}\label{eq:sC}
	{s^C}^* = \frac{N s^C + \alpha \Delta t \mathcal{F}s^F }{ N + \alpha \Delta t \left( \mathcal{F} + \bar{\eta}_N \right)},
\end{equation}
where
\begin{equation}\label{eq:zetaN}
	\mathcal{F} = 
	\zeta N \left[\frac{ F_{3}F_{0} - s^F\left( F_{3}F_{0}-F_{2}F_{1} \right)}{F_{2}F_{1}}\right]^2
\end{equation}

The collisional source term for the number density $N$ in equation~\eqref{eq:evolution_n} is given by
\begin{equation}
	C_{0} = \bar{\eta}_N - \bar{\kappa}_N \frac{J}{\langle \epsilon \rangle} = \bar{\eta}_N - \frac{\bar{\kappa}_N J N }{W \left( E - F_i v^i\right)},
\end{equation}
where $\bar{\eta}_N$ and $\bar{\kappa}_N$ are the energy-integrated number emission and the energy-averaged number absorption, respectively.
This source term is handled in the implicit step.
For instance, in the implicit step, after the neutrino energy and flux density $\left\{E, F_i\right\}$ are updated, we update the neutrino number density $N$ by
\begin{equation}
	{\sqrt{{\gamma}/\hat{\gamma}} N}^* = \frac{\sqrt{{\gamma}/\hat{\gamma}} N + \alpha \sqrt{{\gamma}/\hat{\gamma}} \Delta t \bar{\eta}_N }{ 1 + \alpha \Delta t \bar{\kappa}_N \left( \frac{J}{W\left(E-F_iv^i\right)} \right)}.
\end{equation}

\subsection{Opacities and emissivities}
The energy-averaged absorption and scattering coefficients are
\begin{align}
	\bar{\kappa}_{a,s} = & \frac{\int_0^\infty \kappa_{a,s} \epsilon f \dd{V_\epsilon}}{\int_0^\infty \epsilon f \dd{V_\epsilon}}, \\
	\bar{\kappa}_N = & \frac{\int_0^\infty \kappa_a f \dd{V_\epsilon}}{\int_0^\infty f \dd{V_\epsilon}},
\end{align}
where $f$ is the distribution function, and we have defined $\dd{V_\epsilon} \equiv 4 \pi \epsilon^2 \dd{\epsilon} $.
In this scheme, neutrinos are assumed to follow a Fermi-Dirac distribution function
\begin{equation}
	f_{\rm FD} \left(\epsilon, T_\nu, \mu_\nu \right) = \left[ 1+\exp\left(\frac{\epsilon - \mu_\nu}{k_{\rm B}T_\nu}\right) \right]^{-1},
\end{equation}
where $\epsilon$ is the neutrino energy, $\mu_\nu$ is the neutrino chemical potential, and $T_\nu$ is the neutrino temperature.

In our current implementation, the neutrino chemical potential $\mu_\nu$ can be obtained directly from the equation-of-state table, which assumes that neutrinos are in equilibrium with the fluid.
Namely, we set 
\begin{equation}\label{eq:mu_nu}
	\mu_{\nu} = \mu_{\nu}^{\rm eq}.
\end{equation}
It is suggested that the neutrino chemical potential $\mu_\nu$ can be obtained by interpolating its value at equilibrium and free streaming limit \citep{2010CQGra..27k4103O, 2015PhRvD..91l4021F, 2024PhRvD.109d4012S}, e.g. $\mu_{\nu} = \mu_{\nu}^{\rm eq} \left[ 1 - \exp \left(-\tau\right) \right]$, where $\tau$ is the optical depth.
Since the estimated optical depth $\tau$ via equation~\eqref{eq:zeta} is not very accurate, and this interpolation has a minimal effect in neutron star mergers~\citep{2015PhRvD..91l4021F}, in the current implementation we do not use this interpolated chemical potential.
We note however that this could lead to significant differences in the context of core-collapse supernovae.

Usually, the emissivity, absorption and scattering coefficients are functions of fluid rest mass density $\rho$, temperature $T$ and electron fraction $Y_e$.
When neutrinos are in equilibrium with the fluid, mostly in optically thick regions, the neutrino temperature is roughly the same as the fluid temperature (i.e. $T_\nu \approx T$).
However, the neutrino temperature can be significantly different from the fluid temperature in non optically thick regions.
By following \cite{2016PhRvD..94l3016F}, we first calculate the energy-averaged absorption and scattering opacities by setting $T_\nu = T$
\begin{align}
	\bar{\kappa}_{a,s}^{\rm eq} = & \frac{\int_0^\infty \kappa_{a,s}\left(\rho,T,Y_e\right) \epsilon f_{FD}\left(\epsilon,T,\mu_\nu \right) \dd{V_\epsilon}}{\int_0^\infty \epsilon f_{FD}\left(\epsilon,T,\mu_\nu \right) \dd{V_\epsilon}}, \\
	\bar{\kappa}_N^{\rm eq}     = & \frac{\int_0^\infty \kappa_a    \left(\rho,T,Y_e\right)          f_{FD}\left(\epsilon,T,\mu_\nu \right) \dd{V_\epsilon}}{\int_0^\infty f_{FD}\left(\epsilon,T,\mu_\nu \right) \dd{V_\epsilon}},
\end{align}
and apply the correction
\begin{equation}\label{eq:kappa_T2}
	\bar{\kappa}_{a,s} = \bar{\kappa}_{a,s}^{\rm eq}\left(\rho, T, Y_e \right) \frac{T_\nu^2}{T^2},
\end{equation}
where the neutrino temperature $T_\nu$ can be estimated by 
\begin{equation}\label{eq:tnu}
	T_\nu = \frac{ F_2\left( \frac{\mu_\nu}{k_{\rm B} T} \right) }{ F_3 \left( \frac{\mu_\nu}{k_{\rm B} T} \right)} \langle \epsilon \rangle,
\end{equation}
where $\langle \epsilon \rangle$ is estimated by equation~\eqref{eq:avg_eps}.
In practice, we limit the correction $1 \leq {T_\nu^2}/{T^2} \leq 10$ for more robust simulations.

The energy-integrated emissivities can then be calculated by imposing Kirchhoff’s law
\begin{align}
	\bar{\eta}   = & \bar{\kappa}_a \frac{4 \pi }{\left(hc\right)^3} F_3\left(\frac{\mu_\nu}{k_{\rm B} T}\right) \left(k_{\rm B}T\right)^4, \label{eq:emiss}\\
	\bar{\eta}_N = & \bar{\kappa}_N \frac{4 \pi }{\left(hc\right)^3} F_2\left(\frac{\mu_\nu}{k_{\rm B} T}\right) \left(k_{\rm B}T\right)^3\label{eq:emiss_num}.
\end{align}

For robustness, we consider neutrino-matter interactions only for the region where $\rho \geq 10^{6}~{\rm g/cm^3}$ and $T \geq 0.1~{\rm MeV}$.

\subsection{Coupling to fluid}
The coupling to fluid is the same as described in \cite{2023ApJS..267...38C}, except that the source terms for the evolution equation of the electron fraction $Y_e$ can be obtained from neutrino number densities.
In particular, the evolution equation for the electron fraction $Y_e$ is given by
\begin{equation} 
	\nabla_\alpha\left(\rho Y_e u^\alpha \right) = m_{\rm{u}} \mathcal{R}, 
\end{equation} 
where $m_{\rm{u}}$ is the atomic mass unit, and 
\begin{equation} 
	\mathcal{R} = - \sum_{\nu_i} {\rm sign}\left(\nu_i\right) \left[\bar{\eta}_{N_{\nu_i}}-\frac{\bar{\kappa}_{N_{\nu_i}} J}{W\left(E-F_iv^i\right)} \right],
\end{equation} 
and
\begin{equation} 
	{\rm sign}\left(\nu_i\right) = 
	\begin{cases} 
		+1	,&	\text{if } \nu_i = \nu_e \\
		-1	,&	\text{if } \nu_i = \bar{\nu}_e \\
		0	,&	\text{otherwise } 
	\end{cases} .
\end{equation} 

\subsection{Improvement by \cite{2024AnA...687A..55A}}
The grey scheme of \cite{2016PhRvD..94l3016F} has been improved for core-collapse supernovae by \cite{2024AnA...687A..55A}. These improvements have been partially integrated in \texttt{Gmunu} for more comprehensive comparisons.
Note that, the improvements proposed in \cite{2024AnA...687A..55A} are focused on the core-collapse supernovae problem. Some of them may not apply for neutron star merger cases.
To keep the scheme generic and for easier implementations, we included only the following adjustments.

We replace equation~\eqref{eq:eF}, \eqref{eq:sC}, and \eqref{eq:zetaN} by
\begin{align}
	\frac{\langle \epsilon^F \rangle }{\langle \epsilon \rangle} = &
	\frac{F_{3}F_{1} - s^F\left( F_{3}F_{1}-F_{2}F_{2} \right)}{F_{3}F_{1} - s^C\left( F_{3}F_{1}-F_{2}F_{2} \right)}, \\
	{s^C}^* = & \frac{N s^C + \alpha \Delta t \mathcal{F}s^F }{ N \left(1+\tau\right) + \alpha \Delta t \left( \mathcal{F} + \bar{\eta}_N \right)}, \\
	\mathcal{F} = &
	\zeta N \left[\frac{ F_{3}F_{1} - s^F\left( F_{3}F_{1}-F_{2}F_{2} \right)}{F_{2}F_{2}}\right]^2,
\end{align}
respectively.

When the neutrino temperature $0.1~{\rm MeV} \leq T_\nu \leq 16~{\rm MeV} $, instead of extrapolating the opacities by multiplying the correction factors, we calculate the opacities by 
\begin{align}
	\bar{\kappa}_{a,s} = & \frac{\int_0^\infty \kappa_{a,s}\left(\rho,T,Y_e\right) \epsilon f_{FD}\left(\epsilon,T_\nu,\mu_\nu \right) \dd{V_\epsilon}}{\int_0^\infty \epsilon f_{FD}\left(\epsilon,T_\nu,\mu_\nu \right) \dd{V_\epsilon}}, \\
	\bar{\kappa}_N     = & \frac{\int_0^\infty \kappa_a    \left(\rho,T,Y_e\right)          f_{FD}\left(\epsilon,T_\nu,\mu_\nu \right) \dd{V_\epsilon}}{\int_0^\infty f_{FD}\left(\epsilon,T_\nu,\mu_\nu \right) \dd{V_\epsilon}}.
\end{align}
For the case that the neutrino temperature is higher than 16 MeV, we use the original scheme as in equation~\eqref{eq:kappa_T2}.
Here the neutrino temperature $T_\nu$ is still estimated by equation~\eqref{eq:tnu}.
Although the calculation of the degeneracy parameter should involve neutrino temperature in this case, we do not solve a non-linear equation for neutrino temperature.

Note that, as pointed out in \cite{2024AnA...687A..55A}, estimating the optical depth with equation~\eqref{eq:zeta} not only requires a parameter $\beta$, but also could underestimate the optical depth $\tau$.
However, as the geometry in neutron star mergers is often non-spherical, the alternative approach proposed in \cite{2024AnA...687A..55A} may not work well in such systems.
Additionally, the fact that the optical depth can be calculated from only local variables when using equation~\eqref{eq:zeta} is computationally convenient.
Therefore, we adopt that equation to calculate optical depth $\tau$ in this work.

\section{\label{sec:addon}Add-on implementation}

\subsection{Toward weak equilibrium}
Since the neutrinos-fluid coupling is done explicitly, the hydrodynamical quantities such as temperature $T$ and electron fraction $Y_e$, and therefore the emissivity and opacities, are kept fixed during an implicit time step.
As pointed out in \cite{2022MNRAS.512.1499R}, this can cause the numerical scheme to oscillate if the weak equilibration timescale is too small to be resolved by a timestep $\Delta t$.
In the worst case scenario, this could cause significant changes of the electron fraction $Y_e$ in just one implicit step, affecting the stability of the evolution and leading to failure of the simulation.
This can be avoided if the timestep is chosen to be adaptive to the change of $Y_e$ (e.g. in \cite{2023ApJS..267...38C}).
However, reducing timestep size means increasing the computational cost.
To avoid such strong restrictions of the timestep size without solving the full implicit neutrino-fluid coupling equations, we follow the method proposed by \cite{2022MNRAS.512.1499R}.
In practice, we found that this approach works well in both energy-integrated or dependent neutrino transport simulations.
For completeness, here we highlight the key details of this method.

The idea is to recalculate the emissivity with a new estimated target fluid temperature and electron fraction when the weak equilibration timescale cannot be resolved by a given timestep $\Delta t$.
In particular, when the estimated radiation-fluid equilibration timescale $\tau_{\rm eqm}$ is smaller than $\Delta t$, we estimate the target fluid temperature and electron fraction $\left\{ T^{*}, Y_e^{*}\right\}$ by
\begin{align}
	Y_e^* =& Y_e + \frac{Y_e^{\rm eq} - Y_e}{0.5 \Delta t} \max\left( \tau_{\rm eqm}, 0.5 \Delta t\right), \\
	T^* =& T + \frac{T^{\rm eq} - T}{0.5 \Delta t} \max\left( \tau_{\rm eqm}, 0.5 \Delta t\right),
\end{align} 
where $\left\{ T^{\rm eq}, Y_e^{\rm eq}\right\}$ are the temperature and electron fraction obtained by assuming weak equilibrium and lepton and energy conservation by following \cite{2019EPJA...55..124P}.
After that, we recalculate the emissivity as in equation~\eqref{eq:emiss} and \eqref{eq:emiss_num} with $\left\{ T^{*}, Y_e^{*}\right\}$, while the opacities remain unchanged.

The radiation-fluid equilibration timescale is estimated as
\begin{equation}\label{eq:tau_nu}
	{\tau}_{\rm eqm} = \min_{\nu_i=\nu_e,\bar{\nu}_e; \; \epsilon }\left( \tilde{\tau}_{\rm eqm} \right),
\end{equation} 
where, 
\begin{equation} 
	\tilde{\tau}_{\rm eqm} = \left[ \kappa_a \left( \kappa_a + \kappa_s \right) \right] ^{-1/2}
\end{equation} 
is function of the species and energy of neutrinos.
Since unwanted strong oscillations of the electron fraction $Y_e$ are the main concern, here we consider only the timescale of electron type neutrinos (i.e. $\nu_e$ and $\bar{\nu}_e$).
In the energy-integrated neutrino transport cases, the timescale $\tilde{\tau}_{\rm eqm}$ is calculated directly from the energy-averaged opacities $\bar{\kappa}_{a,s}$; equation~\eqref{eq:tau_nu} in these cases is independent of energy bins.

The fluid temperature and electron fraction $\left\{ T^{\rm eq}, Y_e^{\rm eq}\right\}$ assuming weak equilibrium and lepton and energy conservation, are obtained by solving (see \cite{2019EPJA...55..124P}).
\begin{align}
	Y_l =& Y_e^{\rm eq} + Y_{\nu_e}\left(T^{\rm eq}, Y_e^{\rm eq}\right) - Y_{\bar{\nu}_e}\left(T^{\rm eq}, Y_e^{\rm eq}\right), \label{eq:cons_l} \\
	e_{\rm tot} =& e_{\rm fluid}\left(T^{\rm eq}, Y_e^{\rm eq}\right) + \label{eq:cons_e} \\
	&\frac{\rho}{m_u}\Big[ Z_{\nu_e}\left(T^{\rm eq}, Y_e^{\rm eq}\right) + Z_{\bar{\nu}_e}\left(T^{\rm eq}, Y_e^{\rm eq}\right) + 4Z_{\nu_x}\left(T^{\rm eq}\right) \Big], \nonumber
\end{align} 
where $Y_l$ is the total lepton fraction, $e_{\rm tot}$ and $e_{\rm fluid}$ are the total and fluid energy densities.
The fluid energy density is defined as $e_{\rm fluid} \equiv \rho \left( 1 + \varepsilon \right) $, where $\varepsilon$ is the fluid specific internal energy.
The total lepton fraction $Y_l$ and the total energy density $u$ are obtained at the beginning of the implicit step (the intermediate values of the radiation field is used, after applying the explicit terms; the algorithm thus depends on the exact implicit-explicit scheme adopted) by
\begin{align}
	Y_l =& Y_e + \frac{m_{\rm u}}{\rho}\left( N_{\nu_e} - N_{\bar{\nu}_e} \right), \\
	e_{\rm tot} =& e_{\rm fluid} + \Big[ E_{\nu_e} + E_{\bar{\nu}_e} + 4E_{\nu_x} \Big].
\end{align} 
The neutrino particle and energy fractions $Y_{\nu_i}$ and $Z_{\nu_i}$ in equation~\eqref{eq:cons_l} and \eqref{eq:cons_e} are given by
\begin{align}
	Y_{\nu_i}\left(T, Y_e\right) =& \frac{4 \pi m_{\rm u}}{\rho \left(hc\right)^3}F_2\left(\frac{\mu_{\nu_i}}{k_{\rm B}T}\right) \left(k_{\rm B}T\right)^3 , \\
	Z_{\nu_i}\left(T, Y_e\right) =& \frac{4 \pi m_{\rm u}}{\rho \left(hc\right)^3}F_3\left(\frac{\mu_{\nu_i}}{k_{\rm B}T}\right) \left(k_{\rm B}T\right)^4 .
\end{align} 
The non-linear equations~\eqref{eq:cons_l} and \eqref{eq:cons_e} are solved by using the multidimensional Broyden method described in \cite{press1996numerical}.

Note that, although only electron type neutrinos' equilibrium timescale is used to compare to the simulation timestep (see equation~\eqref{eq:tau_nu}), the energy conservation equation being solved in equation~\eqref{eq:cons_e} assumes heavy lepton neutrinos are also in equilibrium.
In practice, there are regions where the electron type neutrinos are expected to be equilibrium but not for heavy lepton neutrinos.
The method described there may force heavy lepton neutrinos to equilibrium in those regions.
A better solution to overcome the timescale issue will be left as future work.

\subsection{\label{sec:rad_init}Initialisation of the radiation quantities}
By default, we initialise the radiation quantities as follows.
When the absorption opacity is non-zero, we initialise the radiation energy and flux densities $\left\{\mathcal{E},\mathcal{F}_i\right\}$ by assuming the neutrinos are fully trapped.
For instance, we set $\left\{\mathcal{E},\mathcal{F}_i\right\}$ by following equation~(79) and (80) in \cite{2023ApJS..267...38C}, and setting the comoving second moment $\mathcal{H}^\mu = 0$:
\begin{align}
	{\mathcal{E}} = &\frac{{\mathcal{J}}}{3}\left( 4 W^2 - 1 \right), \\
	{\mathcal{F}}_i = & \left( \frac{4}{3}W^2{\mathcal{J}} \right) v_i, 
\end{align}
where the neutrino energy density in comoving frame is assumed to be 
\begin{equation}
	\mathcal{J} = \frac{1}{\left(hc\right)^3} \epsilon f_{\rm FD} \left(\epsilon, T, \mu_\nu \right).
\end{equation}

In the energy-integrated cases, we apply the same equations for $\left\{{E},{F}_i\right\}$, with a energy-integrated neutrino energy density $J = \int_0^{\infty} \mathcal{J} \dd V_{\epsilon}$. 
The neutrino number densities are initialised as
\begin{equation}
	N = \frac{\bar{\eta}_N}{\bar{\kappa}_N} \left( \frac{J}{W\left(E-F_iv^i\right)} \right).
\end{equation}
When the auxiliary scalar $s^C$ is evolved, we set it to be zero everywhere at the beginning of the simulations as in \cite{2016PhRvD..94l3016F}.

\subsection{Avoiding neutrino beam crossing at the pole}
One of the major limitations of the two-moment scheme is its failure to describe crossing radiation beams.
Usually, the closure relations are chosen to be asymptotically correct in the optically thick region, but do not work very well in the free-streaming region \citep{2023LRCA....9....1F}.
As a result, the two-moment approach fails to describe crossing radiation beams (see e.g. \cite{2013MNRAS.429.3533S, 2015PhRvD..91l4021F, 2020MNRAS.495.2285W, 2023LRCA....9....1F}).
This causes huge problem in the polar region in the axisymmetric supramassive/hypermassive neutron star simulations.
The radiation gets trapped and keeps gaining energy in the polar regions due to the geometry of the rotating neutron star, eventually leading to a failure of the simulation.

To avoid this, we enforce the radiation to be outgoing in the optically thin region at each timestep.
For instance, when the flux factor $\zeta \equiv \sqrt{{{H}^\mu {H}_\mu}/{{J}^2}}$ is larger than 0.5, we set
\begin{equation}
	F_i \rightarrow {\rm sign}\left(x^i\right) \times \lvert F_i \rvert .
\end{equation}
In this way, the radiation is enforced to be outgoing and the energy is conserved.

\section{\label{sec:pns_test}Hydrostatic test with a proto-neutron star background}
Here, we follow the test used in \cite{2016PhRvD..94l3016F, 2024AnA...687A..55A}.
In particular, we map the hydrodynamical background of a core-collapse simulations of a 30~${\rm M_{\odot}}$ progenitor.
This profile is identical to the one used in \cite{2024AnA...687A..55A}.
To exclude the Doppler shift and gravitational redshift effects, the fluid velocity is set to zero, and the metric to flat space.
We run the simulations with neutrino evolutions only, and run it till 100 ms to reach to a steady-state solution.

Table~\ref{tab:pns_test} compare the steady-state neutrino signals obtained with different schemes.
Our results agree with \cite{2016PhRvD..94l3016F, 2024AnA...687A..55A} qualitatively.
For instance, all the electron type neutrinos behave similarly but not heavy-lepton neutrinos.
As discussed in the main text, the main reason is that the absorption neutrinosphere is deeper into the star than the scattering neutrinosphere for heavy-lepton neutrinos, while the treatment of the regions between these two neutrinospheres is the major difference between all the schemes we include here.
\begin{table*}[]
\centering
\begin{tabular}{c|cccccc}
\multirow{2}{*}{Schemes}                                & \multicolumn{3}{c}{${\rm [10^{51}~erg/s]}$}   & \multicolumn{3}{c}{${\rm [MeV]}$}                                                                                        \\
                                                        & $L_{\nu_e}$ & $L_{\bar{\nu}_e}$ & $L_{\nu_x}$ & $\langle {\epsilon_{\nu_e}} \rangle$ & $\langle {\epsilon_{\bar{\nu}_e}} \rangle$ & $\langle {\epsilon_{\nu_x}} \rangle$ \\ \hline
Spectral                                                & 99.50       & 31.31             & 143.06      & 10.39                                & 12.28                                      & 19.20                                \\
$\langle \epsilon^F \rangle = \langle \epsilon \rangle$ & 118.61      & 31.37             & 105.96      & 10.96                                & 11.47                                      & 25.01                                \\
\cite{2016PhRvD..94l3016F}, $\beta=6$                   & 117.89      & 34.27             & 124.14      & 10.28                                & 10.74                                      & 22.12                                \\
\cite{2024AnA...687A..55A}, $\beta=6$                   & 117.19      & 33.03             & 101.47      & 10.50                                & 10.65                                      & 20.14                               
\end{tabular}
\caption{\label{tab:pns_test}
	Neutrino luminosities and averaged energies in the hydrostatic simulations with a fixed proto-neutron star background.
	The hydrodynamical background is obtained via a simulation of a 30~${\rm M_{\cdot}}$ progenitor.
	The velocities and gravitational field are set to be zeros to exclude the Doppler shift and gravitational redshift effects.
	These quantities are extracted at $r=500~{\rm km}$ at $t=100~{\rm ms}$.
	Our results agree with \cite{2016PhRvD..94l3016F, 2024AnA...687A..55A} qualitatively.
	For instance, all the electron type neutrinos behave similarly but not heavy-lepton neutrinos.
	This is because the absorption neutrinosphere is deeper into the star than the scattering neutrinosphere for heavy-lepton neutrinos, the treatment in the regions between two neutrinosphere is the major difference between all the schemes we include here.
       }
\end{table*}

\section{\label{sec:ccsn_test}Core collapse of a 20 $\rm{M_{\odot}}$ star in one dimension}
In this section, we present the simulation of the core collapse of a massive star in one dimension with two-moment neutrino transport schemes.
Here, we use the same progenitor and equation-of-state as in \cite{2018JPhG...45j4001O, 2024AnA...687A..55A}.
In particular, the 20 $\rm{M_{\odot}}$ solar metallicity progenitor star of \cite{2007PhR...442..269W}, and the SFHo equation-of-state \cite{2013ApJ...774...17S} is used in this test.

The computational domain covers $[0,10^4]~{\rm km}$ for $r$, with the resolution {$N_r = 128$} and allowing $l_{\max}=10$ mesh levels.
The corresponding finest grid size is $\Delta r \approx 153~{\rm m}$.
The refinement criteria is the same as the core-collapse supernova simulations reported in \cite{2023ApJS..267...38C}.
The neutrino microphysics is provided by \texttt{NuLib}~\citep{2015ApJS..219...24O}, where the neutrino interaction sets are the same as in \cite{2018JPhG...45j4001O}.
The energy space is discretised into 18 groups logarithmically from $1~\rm{MeV}$ to $280~\rm{MeV}$.

Energy-coupled neutrino interactions, such as neutrino-electron inelastic scattering, are not included when an energy-integrated neutrino transport is used.
However, those energy-coupled interactions are essentials for the modelling of the collapsing phase of core-collapse supernovae.
To perform end-to-end core-collapse supernova simulations with energy-integrated neutrino transport, we adopt the parametrised deleptonisation scheme of \cite{2005ApJ...633.1042L}.
Note that, instead of including neutrino pressure by following the description in \cite{2005ApJ...633.1042L}, we switch on the neutrino transport before core-bounce. 
For instance, when the maximum rest mass density $\rho_{\max}$ is above the neutrino trapping density $\rho_{\rm trap} \approx 10^{12}~{\rm g/cm^3}$, we switch on the two-moment neutrino transport and enable the coupling between neutrinos and fluid energies and momentums.
When neutrino transport is switched on, we initialise the radiation quantities as described in section~\ref{sec:rad_init}.
The change of the electron fraction is controlled by this deleptonisation scheme until the proto-neutron star is formed.
i.e., the evolution of the electron fraction $Y_e$ is switched on at core-bounce, which is defined as when the matter entropy per baryon is larger or equals to 3 (i.e. $s \geq 3 \ k_{\rm{B}} / {\rm baryon}$) in the core region ($r \lesssim 30 {\rm \ km}$).

To better compare different transport schemes, we perform all the simulations with identical input microphysics except the reference model.
In particular, we perform simulations with the energy-integrated moment scheme of \cite{2016PhRvD..94l3016F}, \cite{2024AnA...687A..55A}, as well as using the simplified scheme setting $\langle \epsilon \rangle = \langle \epsilon^F \rangle$.
In addition, we perform simulations with energy-dependent neutrino transport with the same deleptonisation scheme and neutrino interactions.
To compare to the conventional core-collapse supernovae simulations, we also include simulations with full energy-dependent neutrino transport with conventional neutrino-matter interactions as in \cite{2018JPhG...45j4001O} as a reference model.

\begin{figure*}
	\centering
	\includegraphics[width=\textwidth, angle=0]{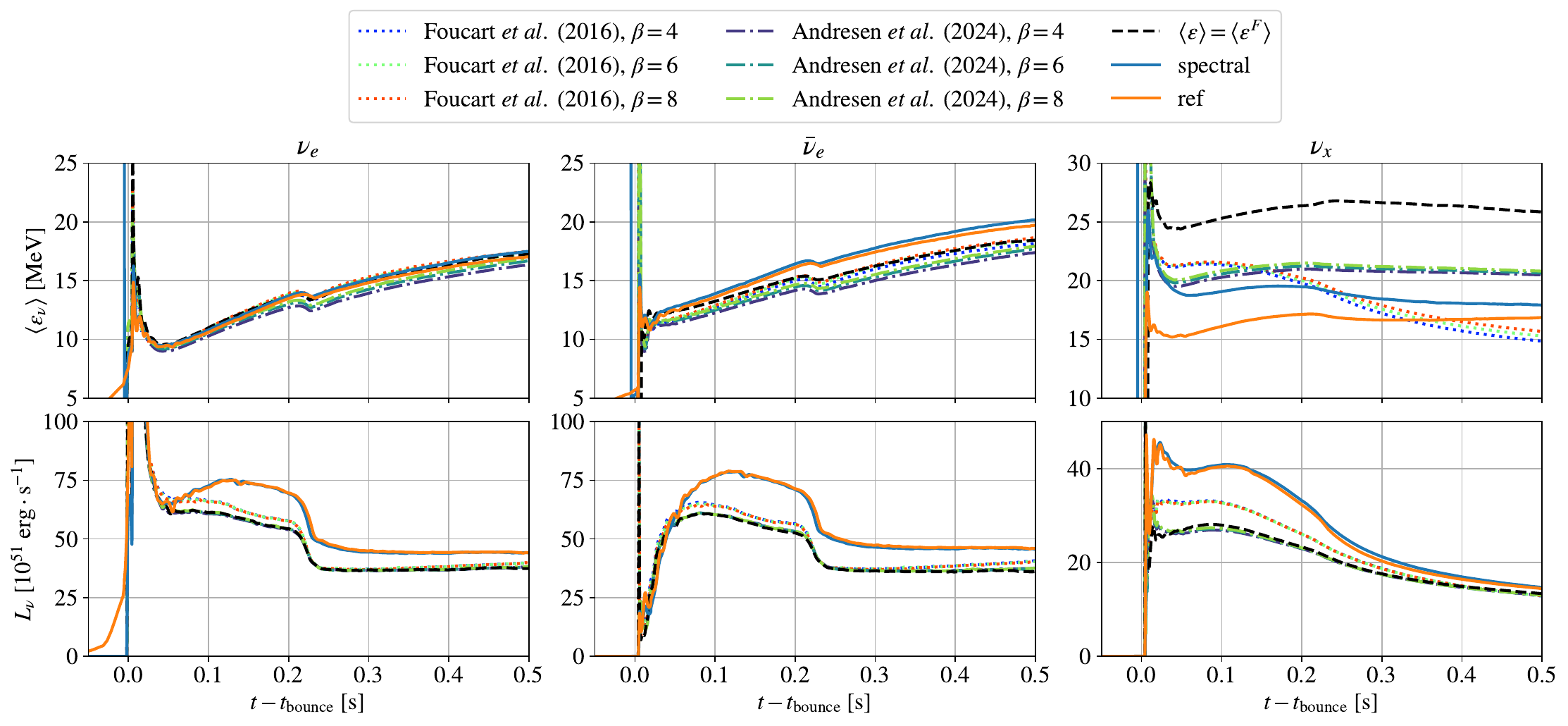}
	\caption{
	Time evolution of far-field averaged neutrino energies ($\langle {\epsilon_\nu} \rangle$, \emph{upper panels}) and luminosities ($L_\nu$, \emph{lower panels}) measured by an observer comoving with fluid at 500 km of a collapsing 20 ${\rm M_{\odot}}$ star.
	The energy-integrated moment scheme of \cite{2016PhRvD..94l3016F} (dotted lines), \cite{2024AnA...687A..55A} (dotted dashed lines) and the simplified version (i.e. setting $\langle \epsilon \rangle = \langle \epsilon^F \rangle$, dashed lines) are shown in the plot.
	The solid blue lines show the energy-dependent neutrino transport simulations with the same neutrino interactions.
	The reference evolution with conventional neutrino-matter interactions (i.e. including neutrino-electron inelastic scattering) are presented with the solid orange lines.
	The averaged neutrino energy of electron types $\langle \epsilon_{\nu_{e}}\rangle$ and $\langle \epsilon_{\bar{\nu}_{e}}\rangle$ are very close among all the cases.
	However, the averaged energy of heavy-lepton neutrinos are very different with different approaches.
	The behaviour of heavy-lepton neutrinos are sensitive to the neutrino treatments, while it is insensitive to the parameter $\beta \in \left[4,8\right]$.
	On the other hand, the luminosities in the case of energy-integrated schemes are lower than the cases with full energy transport, and approaching to similar values at later times.
		}
	\label{fig:ccsn_nu}	
\end{figure*}

The time evolution of averaged neutrino energies $\langle {\epsilon_\nu} \rangle$ measured by an observer comoving with fluid at 500 km of a collapsing 20 ${\rm M_{\odot}}$ star are shown in the left column of figure~\ref{fig:ccsn_nu}.
All the schemes predict very similar averaged neutrino energy of electron types $\langle \epsilon_{\nu_{e}}\rangle$ and $\langle \epsilon_{\bar{\nu}_{e}}\rangle$.
However, the average energies of heavy-lepton neutrinos $\langle \epsilon_{\nu_{x}}\rangle$ are very different with different approaches.
For instance, the simplified approach overestimates by about $\sim 25\%$ the average energy of heavy-lepton neutrinos, when compared to the energy-dependent case.
While the schemes of \cite{2016PhRvD..94l3016F} and \cite{2024AnA...687A..55A} are closer to the energy-dependent case at the very beginning, differences become more significant at later times.
Overall, the behaviour of heavy-lepton neutrinos is sensitive to the neutrino treatments, while it is insensitive to the choice of the parameter $\beta$ in the range of $\left[4,8\right]$.

The time evolution of the luminosities $L_\nu$ measured by an observer comoving with the fluid at 500 km of a collapsing 20 ${\rm M_{\odot}}$ star is shown in the right column of figure~\ref{fig:ccsn_nu}.
Although the average energies of electron type neutrinos are very similar among all the schemes we have tested, the corresponding luminosities can be very different from the energy-dependent cases.
The luminosities in the case of energy-integrated schemes are lower than the cases with full energy transport, and approaching similar values at later times.

\begin{figure}
	\centering
	\includegraphics[width=\columnwidth, angle=0]{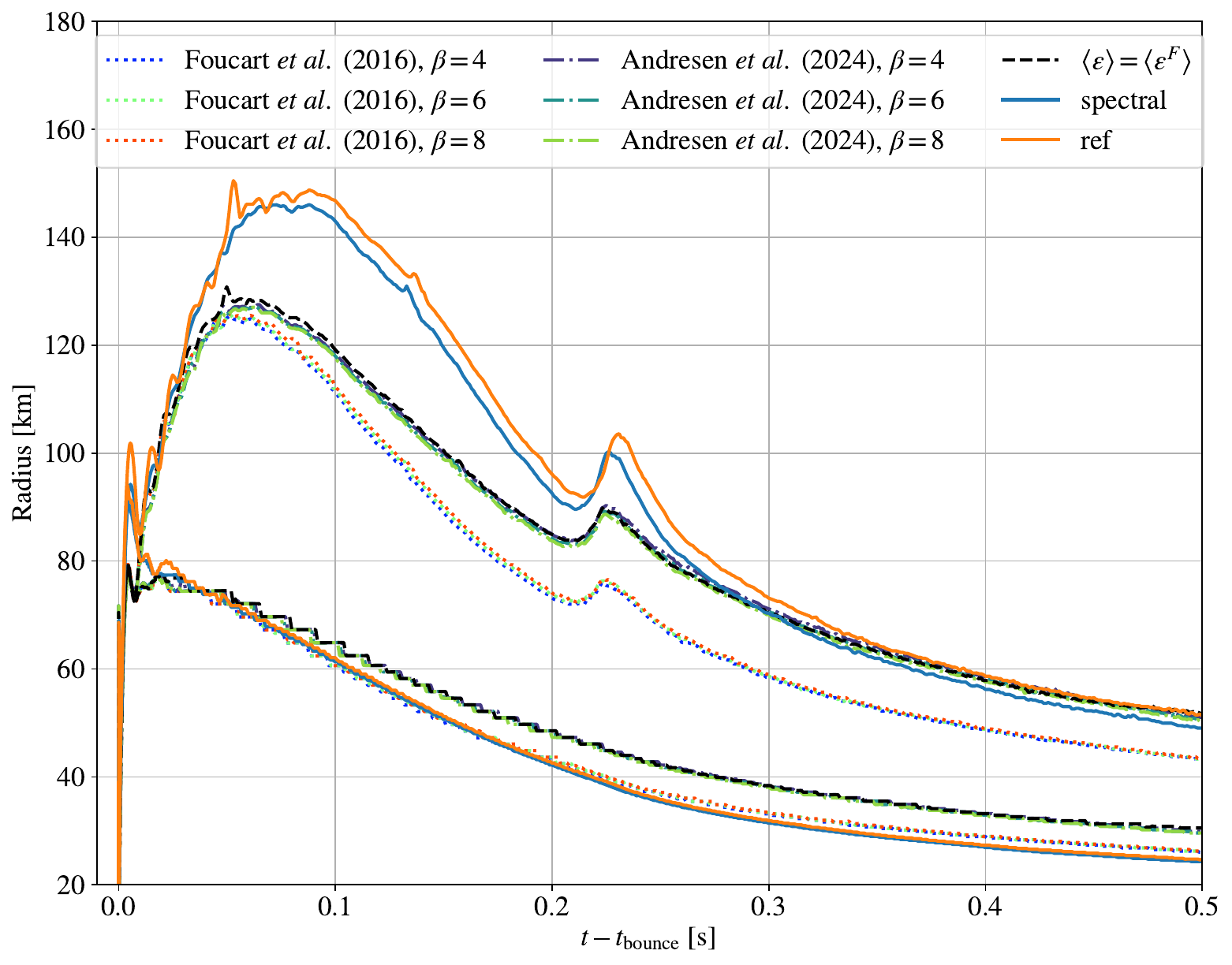}
	\caption{
	Time evolution of the shock and proto-neutron star radius, where the former is defined as the location of the largest absolute velocity while the later is defined as the location when the rest mass density $\rho=10^{11}~{\rm g/cm^3}$.
	The energy-integrated moment scheme of \cite{2016PhRvD..94l3016F} (dotted lines), \cite{2024AnA...687A..55A} (dotted dashed lines) and the simplified version (i.e. setting $\langle \epsilon \rangle = \langle \epsilon^F \rangle$, dashed lines) are shown in the plot.
	The solid blue lines show the energy-dependent neutrino transport simulations with the same neutrino interactions.
	The reference evolution with conventional neutrino-matter interactions (i.e. including neutrino-electron inelastic scattering) are presented with the solid orange lines.
	The shock radius are sensitive to the neutrino treatments but insensitive to the choice of the parameter $\beta \in \left[4,8\right]$.
	The shock radius in the case of energy-integrated schemes are lower than the cases with full energy transport, and approaching to closer values at later times.
	The proto-neutron star radius evolution in the case of \cite{2016PhRvD..94l3016F} matches energy-dependent cases surprisingly well, while other energy-integrated schemes predict slightly larger proto-neutron star.
		}
	\label{fig:ccsn_shock}	
\end{figure}

Figure~\ref{fig:ccsn_shock} compares the time evolution of the shock and proto-neutron star radius among different neutrino transport schemes. 
Here, the shock radius is defined as the location of the largest absolute velocity, while the proto-neutron star radius is defined as the location when the rest mass density $\rho=10^{11}~{\rm g/cm^3}$.
Similar to the neutrino signatures, the shock radius is sensitive to the neutrino treatments but insensitive to the choice of the parameter $\beta \in \left[4,8\right]$.
The shock radius in the case of energy-integrated schemes are lower than the cases with full energy transport, and approaching to closer values at later times.
The proto-neutron star radius evolution in the case of \cite{2016PhRvD..94l3016F} matches energy-dependent cases surprisingly well, while other energy-integrated schemes predict slightly larger proto-neutron star.

Although the energy-integrated transport scheme of \cite{2016PhRvD..94l3016F} implemented in \texttt{Gmunu} is essentially the same as the one reported in \cite{2024AnA...687A..55A}, we report different results.
In particular, we do not observe strong deviations of the average energy of electron type neutrinos compared to energy-dependent scheme, and the average energy of heavy-lepton neutrinos behave differently.
In addition, we observe different hierarchy of the shock radius and proto-neutron star radius.
In our energy-integrated neutrino transport simulations, the shock radius are all smaller than the energy-dependent case, which is the opposite in \cite{2024AnA...687A..55A}.
The proto-neutron star radius of the case of \cite{2016PhRvD..94l3016F} is slightly larger than the energy-dependent case, which is not the case in \cite{2024AnA...687A..55A}. 

Using the equilibrium neutrino chemical potentials everywhere in the simulations (i.e. via equation~\eqref{eq:mu_nu}) is one of the reasons for this discrepancy.
By using an interpolated neutrino chemical potentials with the estimated optical depth, we manage to get results that agree better (e.g. a shock that is stronger than in the energy-dependent case) for the early post-bounce time (i.e. $t - t_{\rm bounce} \lesssim 150~{\rm ms}$), but not afterwards.
Lacking more accurate optical depth information prevents us to investigate this further.
These results suggest that this core-collapse of the 20 $\rm{M_{\odot}}$ star test is very sensitive to the implementation details in energy-integrated cases; comparisons using other codes would be necessary to understand the origin of the discrepancy.


\bibliography{references}{}
\bibliographystyle{aasjournal}



\end{document}